\documentclass[aps,11pt,superscriptaddress]{revtex4-1}
\usepackage{amssymb}
\usepackage{amsmath}
\usepackage{rotating}
\usepackage{ulem}
\usepackage{color}

\begin{document}

\title{How quantizable matter gravitates:\\ a practitioner's guide}
\author{Frederic P. Schuller}\email{fps@aei.mpg.de}\email{while on leave of absence from MPI}
\address{Chair for Quantum Gravity, University of Erlangen-Nuremberg, Staudtstra\ss e\ 7, 91058 Erlangen, Germany\\}
\address{Max Planck Institute for Gravitational Physics\\Am M\"uhlenberg 1, 14476 Potsdam, Germany}
\author{Christof Witte}
\address{Max Planck Institute for Gravitational Physics\\Am M\"uhlenberg 1, 14476 Potsdam, Germany}
\begin{abstract}

We present the practical step-by-step procedure for constructing canonical gravitational dynamics and kinematics directly from any previously specified quantizable classical matter dynamics, and then illustrate the application of this recipe by way of two completely worked case studies. 
Following the same procedure, any phenomenological proposal for fundamental matter dynamics must be supplemented with a suitable gravity theory providing the coefficients and kinematical interpretation of the matter theory, before any of the two theories can be meaningfully compared to experimental data.
\end{abstract}

\date\today
\maketitle
\newpage

\section{Introduction}
There is no reason to assume, and in general it is plainly false, that general relativity still provides a consistent kinematical and dynamical theory of spacetime once the matter fields inhabiting the spacetime are no longer standard model fields. The simple reason for this is that the gravitational dynamics must yield spacetime geometries to which the matter theories at hand can couple without violating elementary physical principles. Indeed, even comparatively innocent-looking 
deviations from the dynamics of standard model matter require an entirely new kinematical and dynamical theory of the underlying spacetime. 

For instance, assume a phenomenologist discovers that some observed spinorial matter field $\Psi$ must be described by a classical field equation of motion of the form, say
\begin{equation*}
\label{DiracEq}
(i \gamma^a + W^a) D_a \Psi = 0\,,
\end{equation*}
which employs a geometric background that features a vector field $W$ in addition to a metric tensor field $g$ (suitably restricted 
such that the spacetime Dirac matrices $\gamma$ and the spin covariant derivative $D$ appearing in the field equation can be constructed). At first sight, such a modification of the Dirac equation indeed seems innocent enough for one to be tempted to stipulate that the dynamics governing the background be still provided by Einstein's gravitational field equations for $g$ and maybe some abelian gauge field dynamics for the vector field $W$. However, we will see that this particular choice of gravitational dynamics would have solutions that render the above matter theory either non-predictive (thus not even classically acceptable), non-quantizable, or both. With predictivity being an unconditional feature of any classical matter theory and quantizability ensuring relevance beyond the classical domain, this result is clearly unacceptable. One may thus either reject the above matter field dynamics as unphysical, or, if our phenomenologist insists that this equation describes observable fundamental matter, we must instead provide another gravity theory whose solutions render the matter theory predictive and quantizable. Are there such gravitational dynamics that can underpin the viciously modified Dirac equation above?

This question has an intriguing---and even constructive---answer. Not only for the above example, but indeed for {\it any} specific linear matter dynamics, one can derive the complete kinematical and dynamical contents of the underpinning gravity theory directly from the matter field equations it is supposed to carry; for the technical derivation see \cite{RRS,GSWW}. The only construction principle is that the resulting gravitational kinematics and dynamics must render the assumed matter field equations both predictive and quantizable; everything else  follows from mathematical theorems. 

Showing that these two basic assumptions already completely fix the {\it kinematics}---such as the distinction of initial data surfaces, the construction of observer frames and thus the interpretation of matter field components, massive and massless dispersion relations, the duality maps associating momenta and velocities for massive and massless particles, and so forth---requires the employment of an intricate interplay of real algebraic geometry, convex analysis and the theory of partial differential equations \cite{RiveraPhD}. The central result is that in order to enable predictivity, the principal polynomial of the matter field equations must be hyperbolic, and in order to enable quantizability, the associated dual polynomial must be hyperbolic as well. This bi-hyperbolicity imposes so severe a constraint on the coefficients featuring in the matter field equations that the above kinematical constructions are uniquely fixed.

With the kinematical structure of the theory determined, the coefficients featuring in the matter field equations must then follow {\it dynamics} whose initial-value formulation is commensurate with the kinematically determined projection of the spacetime geometry to initial data surfaces. In other words, the dynamics must be such that it evolves geometric initial data between hypersurfaces that also serve as initial data surfaces for the given matter field equations. Casting this idea into tractable mathematical form, one proceeds principally along the same lines that were laid out four decades ago by geometrodynamicists \cite{HKT,Kuchar:1974es}, but with the technical scope vastly extended to any bi-hyperbolic spacetime geometry. The final result of this effort, derived in \cite{GSWW} and explained in great conceptual and technical detail in \cite{WittePhD}, are the master equations reproduced on page \pageref{mastergleichungen} of the present paper. The master equations are a set of linear homogeneous partial differential equations, whose coefficients are constructed directly from the coefficients featuring in the specified matter field equations and whose solution provides (the collection of coefficients of a series expansion of) the gravitational Lagrangian.       


The present paper is concerned with cutting away the heavy technical baggage that comes with the derivation of the above results, and instead manages to condense their practical implications into an easily executable recipe, by which one constructs the master equations from any given linear matter field dynamics in eight easy steps. The relevance of the such constructed master equations is that
\begin{center}
{\it A solution to the master equations is \\
a gravity theory that can carry the specified matter dynamics.} 
\end{center}
Thus the master equations must be practically solved, in a ninth step, in order to obtain a concrete gravitational Lagrangian. In cases where such a solution of the master equations is difficult to obtain, one may inject at this stage, as a tenth step, additional physical assumptions such as energy conditions on the matter or (compact) symmetry assumptions on the spacetime geometry in order to simplify the master equations. Such additional assumptions, however, are not fundamentally needed and the master equations are already uniquely determined without them. Any additional assumptions beyond predictivity and quantizability only serve as a possibly convenient means to the end of extracting information from the full master equations, for specific physical situations where the master equations simplify to a more tractable form. 

Two completely worked case studies---namely the comparatively simple derivativation of the Einstein-Hilbert Lagrangian as the unique solution to the master equations determined by Maxwell electrodynamics in section \ref{sec:caseI}, on the one hand, and the more involved derivation of the gravitational dynamics that underlie some prototypical non-standard model matter dynamics in section  \ref{sec:caseII} on the other hand---present illustrations of the general ten-step procedure described in section \ref{sec:recipe}. These case studies are indeed illustrate both the technicalities of the recipe and its significance in three respects. First, they are an instance of the rule that an example sometimes says more than a thousand words; having worked through the two case studies, the reader will have no difficulty in applying the recipe to the matter model of his interest. Secondly, the first case study reveals that the complete kinematics and dynamics of general relativity are simply a consequence of having predictive and quantizable Maxwell (or other standard model) matter dynamics, while the second case study presents an explicit example of non-standard model matter dynamics that are rendered predictive and quantizable only if the underlying gravity is the one derived according to the recipe summarized in this paper.   

The revelance of the simple procedure described and illustrated in this paper---namely for deriving gravitational Lagrangians directly from the dynamics of matter populating the spacetime---of course lies beyond the two specific examples provided here. For it allows to derive a suitable gravity theory for {\it any} matter theory that one may be prompted to consider for phenomenological or theoretical reasons. But this possibility immediately implies an imperative: gravitational kinematics and dynamics must never be postulated, since unless they accidentally coincide with the results of the procedure described in this paper, any such postulates would generically be in contradiction to the quantizability of the matter equations the resulting spacetime geometries must carry. That, conversely, the gravity can instead be fully and quite easily constructed from this consistency postulate is, of course, very good news.
 
\newpage
\section{Practical guide to the derivation of gravity actions}\label{sec:recipe}
The following ten-step procedure provides the simple practical recipe for the construction of 
canonical gravitational dynamics from any previously specified quantizable classical matter dynamics. These rules follow from the results obtained in \cite{RRS} and \cite{GSWW} and can be laid down without any recourse to the heavy technical machinery that was needed for their derivation. To see the abstract rules at work, the reader finds an illustration for each of the steps described here in the two completely worked case studies provided in sections \ref{sec:caseI} and \ref{sec:caseII}.   
\begin{description}
\item[Step 1. Specify test matter dynamics] Provide classical dynamics for a `matter' field $\Phi$ (or a collection of such) on a smooth manifold $M$, by specifying partial differential equations of motion whose coefficients are completely determined by some `geometry' (described by a tensor field $G$ of a priori arbitrary type, or a collection of such), wherein the matter field $\Phi$ takes values in some representation vector space $V$ of the general linear group $GL(\dim M,\mathbb{R})$ (or that of a group defined with recourse to $G$, see the second case study).
Irrespective of any chosen type of matter field or geometry, general coordinate covariance of the matter field equations can be ensured by deriving them from a scalar action functional
$$S_{\textrm{\small matter}}[\Phi,G]$$
by way of variation with respect to the matter field, which will result in field equations valued in the dual space $V^*$. 

Test matter, in particular, is defined by any equation of motion (i) which is linear in the matter field, i.e., takes the form
$$\sum_{n=0}^N Q_{AB}^{a_1 \dots a_n} \partial_{a_1} \cdots \partial_{a_n} \Phi^B = 0\,,$$
where $A, B = 1, \dots, \dim V$ and $\Phi^A$ are the components of the matter field with respect to some basis of the representation space $V$---where the linearity ensures that every solution can be scaled to arbitrarily small amplitudes in order to reduce back-reaction below any desired bound---and (ii) whose coefficients $Q_{AB}^{a_1 \dots a_N}$ of the highest order derivative term are a function of the geometric tensor field $G$ (but not of any of its derivatives)---which ensures that the causal structure of the matter field dynamics is encoded in the spacetime geometry at each point, see the next step. \\

\item[Step 2. Calculate the principal tensor field] If the matter field equations feature no gauge ambiguity---meaning that all components of the tensor field $\Phi$ are uniquely determined by a solution of the field equations---then the principal tensor associated with these field equations is the totally symmetric contravariant tensor field which is constructed from the highest order coefficients $Q^{a_1 \dots a_N}$ of the $N$-th order field equations by virtue of letting  
$$P(k) := \pm \omega \det_{A,B}\left[{Q^{a_1 \dots a_N}_{AB} k_{a_1} \cdots k_{a_N}}\right]\qquad\textrm{(cancel repeated factors)}$$
for every covector field $k$, and where the instruction to cancel repeated factors refers to not further reducible factors whose product $P_{(1)}(k) \cdots P_{(f)}(k) = P(k)$.
If the field equations do contain a gauge ambiguity, first fix the latter by either imposing an explicit gauge condition or transferring to gauge-independent variables. The rank of the totally symmetric tensor $P$ that results from polarization from the above definition will appear explicitly in a number of places and be denoted $\deg P$ throughout. The above construction of the principal tensor is unique up to choice of a scalar density $\omega$ of the appropriate weight in order to render the $P$ a tensor and an overall sign $\pm$ to be chosen later. The choice of density amounts to a choice of volume on the spacetime and would have been used already in the formulation of the matter action if the field equations have been derived from such. 

\item[Step 3. Calculate the dual tensor field] Let $P_{(1)}, \dots, P_{(f)}$ be the mutually distinct irreducible factors (i.e., tensors that themselves cannot be written as the tensor product of two tensors of non-vanishing rank) of the principal tensor field $P$ and
consider for each such $P_{(i)}$ the map $DP_{(i)}$ that maps every covector field $k$ with $P_{(i)}(k)=0$ to the vector field with components
$$(DP_{(i)}(k))^a := (\deg P_{(i)}) P_{(i)}^{a\, a_2 \dots a_{\deg P_{(i)}}}  k_{a_2} \dots k_{a_N}\,,$$
where $\deg P_{(i)}$ denotes the rank of the irreducible factor field $P_{(i)}$. The field $P_{(i)}^\#$ dual to the factor field $P_{(i)}$ is then the totally symmetric contravariant tensor field of lowest rank $\deg P_{(i)}^\#$ (which may differ from $\deg P_{(i)}$) defined by the condition to vanish precisely on the images of the $P_{(i)}$-null covectors,
$$P_{(i)}^\#( DP_{(i)}(k) ) = 0 \quad \textrm{precisely for all } k \textrm{ with } P_{(i)}(k)=0\,.$$ 
The dual tensor field is then defined as the product of the duals of all the irreducible factors,
$$P^\#(X) := P_{(1)}^\#(X) \cdots P_{(f)}^\#(X)$$
for all vector fields $X$, and thus satisfies the duality condition $P^\#(DP(k))=0$ for all $k$ that are $P$-null. The dual tensor always exists (if the tensor field $P$ is hyperbolic, see the next step) and can be constructively obtained by Buchberger's algorithm \cite{Hassett}, which however quickly becomes expensive with increasing rank $\deg P$ of the principal tensor field.


\item[Step 4. Restrict to bi-hyperbolic geometries] 
A necessary condition for the matter equations of motion to be predictive is that the principal tensor field $P$ is hyperbolic \cite{Garding}. This amounts to the simple algebraic condition that there exists an  covector field $h$ such that (i) $P(h)$ is an everywhere non-vanishing function and (ii) for every covector field $q$ the equation 
$$P(h + \lambda q) = 0$$
admits only everywhere real-valued functions $\lambda$ as solutions. Any covector field $h$ with this property is called a hyperbolic covector field. 

A necessary condition that the matter equations be canonically quantizable is that the dual tensor field $P^\#$ is hyperbolic \cite{RiveraPhD}, where hyperbolicity is defined exactly as above, but now with vector fields $H$ and $Q$ taking the role previously played by the covectors fields $h$ and $q$. Any vector field with that property is called a hyperbolic vector field. The overall sign of $P$ can then always be chosen such that every hyperbolic covector field $h$ is $P$-positive, i.e., $P(h)>0$, and we choose to impose this sign convention for definiteness.

Since both the principal and the dual tensor field are defined in terms of the tensor field $G$ providing the spacetime geometry, the hyperbolicity of the former two tensor fields imposes corresponding algebraic conditions on the latter, which immediately exclude certain algebraic classes of geometries. 


\item[Step 5. Determine the geometric degrees of freedom]
While suitable initial data surfaces do not need to be constructed explicitly in order to derive the gravitational dynamics, we assume that such an embedded initial data surface has been chosen and gives rise to linearly independent vector fields $e_1, \dots, e_{\dim M-1}$ along the hypersurface that are tangent to it as well as a covector field $n$ along the hypersurface that annihilates each of the said tangent vector fields and that is hyperbolic (see the previous step) and normalized in the sense that $P(n)=1$.    

Then bases for all spacetime tangent and cotangent spaces along the initial data hypersurface $X(\Sigma)$ are provided by 
$$e_0 := \frac{DP(n)}{\deg P},  e_1, \dots, e_{\dim M-1}\qquad\textrm{and}\qquad\epsilon^0 := n, \epsilon^1, \dots, \epsilon^{\dim M -1}\,,$$
respectively, satisfying the usual duality condition $\epsilon^a(e_b)=\delta^a_b$. Note that the principal tensor field $P$ thus enters explicitly into the definition of $e_0$ and thus implicitly into that of the $\epsilon^1, \dots, \epsilon^{\dim M -1}$. 

Now consider a collection of hypersurface fields $G^{\hat A}$, where the hatted index $\hat A$ runs over all hypersurface index combinations that are required to reconstruct the geometric tensor field $G$ everywhere along the initial data hypersurface from the $G^{\hat A}$ and the above-listed bases. For instance, if $G$ is a $(1,1)$-tensor field, then
\begin{eqnarray}
G &=& G(\epsilon^a,e_b)\, e_a \otimes \epsilon^b\nonumber\\
 &=& \underbrace{G(\epsilon^0 ,e_0)}_{=:\, G^0{}_0} e_0 \otimes \epsilon^0  + \underbrace{G(\epsilon^0, e_\beta)}_{=: \,G^{0}{}_\beta} e_0 \otimes \epsilon^\beta + \underbrace{G(\epsilon^\alpha,e_0)}_{=:\,G^\alpha{}_0} e_\alpha \otimes \epsilon^0 + \underbrace{G(\epsilon^\alpha,e_\beta)}_{=:\, G^\alpha{}_\beta} e_\alpha\otimes \epsilon^\beta\nonumber
\end{eqnarray}
and thus $G^{\hat A} = (G^0{}_0, G^0{}_\beta, G^\alpha{}_0, G^\alpha{}_\beta )$ consists of one hypersurface scalar, one hypersurface covector, one hypersurface vector and one hypersurface endomorphism field. The hatted index $\hat A$ would thus range, in this case, over the values
$$\hat A \quad \in \quad \left\{{}^0{}_0\,, \,\,{}^0{}_\beta\,, \,\,{}^\alpha{}_0\,,\,\, {}^\alpha{}_\beta\right\}\,.$$
For any other valence of the geometric tensor field $G$, one proceeds in exactly analogous fashion. 

But now since the hypersurface fields $G^{\hat A}$ determine the geometric tensor field $G$, which in turn determines the principal tensor field $P$, the above duality conditions between the tangent and cotangent space bases amount to precisely $\dim M$ conditions
$$P(\epsilon_0) = 1 \qquad \textrm{ and } \qquad L(\epsilon^0)(\epsilon_\alpha)= 0$$
relating the hypersurface fields $G^{\hat A}$. 

Thus only an unconstrained subset $G^A$ (for a suitable range of the unhatted index $A$) of the above hypersurface fields $\hat G^{\hat A}$, whose choice automatically implements the above conditions, presents independent geometric degrees of freedom (see, for instance, the first case study). However, in some cases it may be convenient or even necessary to make suitable field redefinitions at this point in order to find a workable set of unconstrained degrees of freedom (see, for instance, our second case study). 


\item[Step 6. Calculate the coefficients of the master equations]
For each independent geometric hypersurface field $G^A$ that has been obtained directly by projection of the spacetime geometry $G$ as described in the previous step, construct the coefficient functions
\begin{equation}
M^{A \gamma} := \left\{
\begin{array}{lll}
\textrm{for each } \quad G^{\dots 0 \dots}{}_{\dots}  \quad&\textrm{ include a summand }& \quad  - G^{\dots \gamma \dots}{}_{\dots}\\
\textrm{for each } \quad G^{\dots}{}_{\dots 0 \dots }  \quad&\textrm{ include a summand }& \quad  -(\deg P\!-\!1) \,G^{\dots}{}_{\dots \alpha \dots} P^{\alpha\gamma}\\
\textrm{for each } \quad G^{\dots \alpha \dots}{}_{\dots}  \quad&\textrm{ include a summand }& \quad  (\deg P\!-\!1) \, G^{\dots 0 \dots}{}_{\dots} P^{\alpha\gamma}\\
\textrm{for each } \quad G^{\dots}{}_{\dots \alpha \dots}  \quad&\textrm{ include a summand }& \quad  - G^{\dots}{}_{\dots 0 \dots} \delta^\gamma_\alpha
\end{array}\right.\nonumber
\end{equation}
where the dots represent indices that are kept unchanged, and similarly,
\begin{equation}
U^{A\,\rho\chi} := \left\{
\begin{array}{lll}
\textrm{for each } \quad G^{\dots \alpha \dots}{}_{\dots}  \quad&\textrm{ include a summand }& \quad  - P^{\chi \alpha} G^{\dots \rho \dots}{}_{\dots}\\
\textrm{for each } \quad G^{\dots}{}_{\dots \alpha \dots}  \quad&\textrm{ include a summand }& \quad  P^{\chi\xi} \delta^\rho_\alpha G^{\dots}{}_{\dots \xi \dots}
\end{array}\right.\nonumber
\end{equation}
as well as
\begin{equation}
V^{A\,\chi} := P^{\chi\xi} \partial_\xi G^A + \left\{
\begin{array}{lll}
\textrm{for each } \quad G^{\dots \alpha \dots}{}_{\dots}  \quad&\textrm{ include a summand }& \quad  P^{\chi\alpha} \partial_\lambda G^{\dots \lambda \dots}{}_{\dots}\\
\textrm{for each } \quad G^{\dots}{}_{\dots \alpha \dots}  \quad&\textrm{ include a summand }& \quad  -P^{\chi\lambda} \partial_\alpha G^{\dots}{}_{\dots \lambda \dots}
\end{array}\right.\nonumber
\end{equation}
Note that in case some field redefinitions have been performed {\it after} projecting the spacetime geometry $G$ to the hypersurface, the redefined fields will be some function of the originial projections, and in this case, the coefficients $M^{A\, \gamma}$, $U^{A\,\rho\chi}$ and $V^{A\,\chi}$ associated with the redefined fields are to be calculated from the respective coefficients associated with the original projected fields by virtue of product and chain rules (for an illustration, see the second case study).
 
Finally, in terms of the above coefficient functions calculate \\[-24pt]
\begin{eqnarray*}
Q_A{}^{B\,\gamma} &:=& -\frac{\partial M^{B\,\gamma}}{\partial \hat G^A}\,,\nonumber\\
T^{A\,[\mu\nu]} &:=& -\,Q_B{}^{A\,[\mu}M^{|B|\,\nu]}+ U^{A\,[\mu\nu]}\,,\nonumber\\
S^{A\,\gamma} &:=& \partial_\beta Q_B^{~~A(\beta|}M^{B|\gamma)}-Q_{B}^{~~A[\beta|}\partial_\beta M^{B|\gamma]}-\partial_\beta U^{A\,(\beta\gamma)}-V^{A\,\gamma}\,,\nonumber\\[-36pt]
\end{eqnarray*}
which completes the calculation of all coefficients needed to set up the master equations.


\item[Step 7: Set up the master equations] The coefficient functions calculated in Step 6 already completely determine the gravitational master equations displayed on the next page. The master equations are equations for the weight-one tensor densities 
$$C=C(G^A,\partial G^A, \partial\partial G^A,\partial\partial\partial G^A) \qquad\textrm{ and } \qquad C_{B_1\dots B_N}=C_{B_1 \dots B_N}(G^A,\partial G^A, \partial\partial G^A)\,,$$
to which we will refer as the `scalar potential' and the `tensor potentials', respectively. 

\begin{sidewaystable}\label{mastergleichungen}
    \centering
    \begin{tabular}{ll}
    & \hspace{5cm}MASTER EQUATIONS DETERMINING THE GRAVITATIONAL LAGRANGIAN\\[12pt]
   &\hspace{1cm} for the weight-one tensor densities \quad $C=(\hat G^A, \partial \hat G^A, \partial\partial\hat G^A, \partial\partial\partial \hat G^A)
  \quad \textrm{and} \quad C_{B_1\dots B_{N\geq1}}=(\hat G^A, \partial \hat G^A, \partial\partial\hat G^A)$ \quad are the \\[24pt] 
       & \hspace{10cm} SIX EQUATIONS\\
   (1) & $0=\frac{\partial C_{B_1}} {\partial\,\partial^2_{(\beta_1\beta_2|}\hat G^A}\,M^{A\,|\beta_3)}+\frac{\partial C}{\partial\,\partial^3_{\beta_1\beta_2\beta_3} \hat G^{B_1}}$\\
 (2) & $ 0 =  2\, C_{AB_1} U^{A\,(\alpha\beta) } - \, \frac{\partial C_{B_1}}{\partial\partial_{(\beta|} \hat{G}^A} M^{A\,|\alpha)} - 2 \, \frac{\partial C_{B_1}}{\partial\partial^2_{(\beta|\gamma}\hat{G}^A} \partial_\gamma M^{A\,|\alpha)} +\,\frac{\partial C}{\partial\partial^2_{\alpha\beta} \hat{G}^{B_N}}-3\,\partial_\gamma\frac{\partial C}{\partial\,\partial^3_{\alpha\beta\gamma} \hat G^{B_1}}$\\
 (3) & $ 0 = 2\, C_{A B_1}  (S^{A\,\alpha}+2\,\partial_\mu T^{A\,[\mu\alpha]})+2\,\partial_\mu C_{AB_1 }T^{A\,[\mu\alpha]} - \,Q_{B_1}{}^{M\,\alpha}\,C_{M}
          + \, \frac{C_{B_1}}{\partial \hat{G}^A} M^{A\, \alpha}+ \, \frac{\partial C_{B_1}}{\partial\partial_{\gamma} \hat{G}^A} \partial_\gamma M^{A\,\alpha} 
       + \, \frac{\partial C_{B_1}}{\partial\partial^2_{\gamma\delta} \hat{G}^{A}} \partial^2_{\gamma\delta} M^{A\,\alpha} +\frac{\partial C}{\partial\partial_{\alpha} \hat{G}^{B_1}} 
           - 2\, \partial_\gamma  \frac{\partial C}{\partial\partial^2_{\alpha\gamma} \hat{G}^{B_1}}+3\,\partial^2_{\beta\gamma}\frac{\partial C}{\partial\, \partial^3_{\alpha\beta\gamma} \hat G^{B_1}}$\quad$ $\\
 (4) & $0 = 2 \partial_\mu(C_A U^{A\,(\beta\mu)}) +2 C_A S^{A\,\beta}+2\partial_\nu C_A T^{A\,[\nu\beta]}+ 2 \frac{\partial C}{\partial \hat{G}^A} \,M^{A\,\beta}+ 2 \frac{\partial C}{\partial\partial_{\mu}\hat{G}^A} \,\partial_\mu M^{A\,\beta}+ 2  \frac{\partial C}{\partial\partial^2_{\mu\nu}\hat{G}^A}\,\partial^2_{\mu\nu} M^{A\,\beta}+ 2  \frac{\partial C}{\partial\partial^3_{\mu\nu\rho}\hat{G}^A}\,\partial^2_{\mu\nu\rho} M^{A\,\beta} $\\
& \qquad$-  \partial_\mu\left(2\frac{\partial C}{\partial\partial_{(\mu|}\hat{G}^A} M^{A\,|\beta)}+ 4 \frac{\partial C}{\partial\partial^2_{(\mu|\nu} \hat{G}^A} \partial_{\nu}M^{A\,|\beta)}+6\frac{\partial C}{\partial\partial^3_{(\mu|\nu\rho} \hat{G}^A} \partial^2_{\nu\rho}M^{A\,|\beta)}\right)+\partial^2_{\mu\nu}\left(3 \frac{\partial C}{\partial \,\partial^2_{(\mu\nu|}\hat G^A}M^{A\,|\beta)}+ 9\frac{\partial C}{\partial \,\partial^2_{(\mu\nu|\rho}\hat G^A}\partial_\rho M^{A\,|\beta)}\right)- 4 \partial^3_{\mu\nu\rho}\left(\frac{\partial C}{\partial\,\partial^3_{(\mu\nu\rho|} \hat G^A} M^{A\,|\beta)}\right)$\\
 (5) & $0=\partial_\alpha\left(\frac{\partial C}{\partial\,\partial^2_{(\beta_1|\alpha}\hat G^A}\,M^{A\,|\beta_2)}+4\, \frac{\partial C}{\partial\,\partial^3_{(\beta_1|\alpha\gamma}\hat G^A }\,\partial_\gamma M^{A\,|\beta_2)}-2\partial_\delta\Big\{\frac{\partial C}{\partial\,\partial^3_{\alpha\delta(\beta_1} \hat G^{A}}\,M^{A\,|\beta_2)}\Big\}\right)$\\
 (6) & $0=2\,\frac{\partial C}{\partial\,\partial^2_{(\beta_1\beta_2|}\hat G^A}\,M^{A\,|\beta_3)}+6\, \frac{\partial C}{\partial\,\partial^3_{(\beta_1 \beta_2|\gamma}\hat G^A }\,\partial_\gamma M^{A\,|\beta_3)}-4\,\partial_\gamma\left(\frac{\partial C}{\partial\,\partial^3_{(\beta_1\beta_2|\gamma}\hat G^A}\,M^{A\,|\beta_3)}\right)$\\[24pt]
  & \hspace{9cm} FIVE SEQUENCES ($N\geq 2$)\\
    ($7_N$)  & $0 = \frac{\partial C_{B_1 \dots B_N}}{\partial\partial^2_{(\alpha\beta|} \hat{G}^A} M^{A\,|\gamma)}$\\
    ($8_N$) & $0 = C_{AB_1\dots B_{N}} T^{A\,[\mu\nu]}$\\
    ($9_N$) & $0 = \frac{\partial C_{B_1 \dots \widetilde{B_i} \dots B_N}}{\partial \partial^2_{\mu\nu}\hat{G}^{B_a}} - \frac{\partial C_{B_1 \dots \dots B_{N-1}}}{\partial \partial^2_{\mu\nu}\hat{G}^{B_N}}$\\     
($10_N$)  &  $0 =  (N+1)!\, C_{AB_1\dots B_N} U^{A\,(\alpha\beta) } - N!\, \frac{\partial C_{B_1\dots B_N}}{\partial\partial_{(\beta|} \hat{G}^A} M^{A\,|\alpha)} - 2 N!\, \frac{\partial C_{B_1\dots B_N}}{\partial\partial^2_{(\beta|\gamma}\hat{G}^A} \partial_\gamma M^{A\,|\alpha)}  -(N-2)(N-1)!\, \frac{\partial C_{B_1 \dots B_{N-1}}}{\partial \partial^2_{\alpha\beta} G^{B_N}}$\\
    ($11_N$) \quad\quad & $0 = (N+1)! C_{A B_1 \dots B_N}  (S^{A\,\alpha}+2\,\partial_\mu T^{A\,[\mu\alpha]})
        +(N+1)!\,\partial_\mu C_{AB_1 \dots B_N}T^{A\,[\mu\alpha]} - NN!\,Q_{(B_1}{}^{M\,\alpha}\,C_{B_2\dots B_N)M}$\\
 &\qquad  $+ N!\, \frac{C_{B_1\dots B_N}}{\partial \hat{G}^A} M^{A\, \alpha}+ N!\, \frac{\partial C_{B_1\dots B_N}}{\partial\partial_{\gamma} \hat{G}^A} \partial_\gamma M^{A\,\alpha} 
       + N!\, \frac{\partial C_{B_1 \dots B_N}}{\partial\partial^2_{\gamma\delta} \hat{G}^{A}} \partial^2_{\gamma\delta} M^{A\,\alpha}+ (N-1)! \sum_{a=1}^N \frac{\partial C_{B_1 \dots \widetilde{B_a} \dots B_N}}{\partial\partial_{\alpha} \hat{G}^{B_a}} 
           - 2(N-1)! \partial_\gamma  \frac{\partial C_{B_1 \dots B_{N-1}}}{\partial\partial^2_{\alpha\gamma} \hat{G}^{B_N}}$\\[24pt]
 & whose coefficient functions $U^{A\mu\nu}, V^{A\gamma}, M^{A\gamma}, Q_A{}^{B\gamma}, T^{A\mu\nu}$ and $S^{A\gamma}$ are determined by the matter action $S[\Phi,G]$, according to Steps 1 to 6.
\end{tabular}
\end{sidewaystable}

The significance of these potentials is that they completely define the gravitational Lagrangian density
$$L[G](K) = \sum_{N=0}^\infty C_{B_1 \dots B_N}[G] K^{B_1} \dots K^{B_N}$$
in terms of the geometric hypersurface fields $G^A$ and their velocities $K^A$, such that the Euler-Lagrange equations for the geometry are 
$$\frac{\partial}{\partial t} \left(\frac{\partial L(z)}{\partial K^A(z)}\right)=\int_\Sigma\, dx\,\left[N(x)\frac{\delta L(x)}{\delta G^A(z)}\right]+\mathcal L_{\vec N}\left(\frac{\partial L(z)}{\partial K^B(z)}\right)+\partial_\beta\left(N(z)\, Q_A{}^{B\,\beta}(z)\right)\,\frac{\delta L(z)}{\delta K^B(z)}\,,
$$
where the integral is over the hypersurface, supplemented by the kinematical relation 
$$\dot{G}^A(z)=N(z)  K^A(z)+\partial_\gamma N(z)\, M^{A\,\gamma}(z) + \mathcal L_{\vec N} G^A(z)\,,$$
where $N$ is a freely specifiable lapse function and $\vec N$ is a freely specifiable shift vector field on the initial data hypersurface. 
Note that while the scalar density $C$ may depend on up to third derivatives of the geometric tensor fields, the tensor densities $C_{B_1 \dots B_N}$ depend on at most second derivatives. 

\item[Step 8. Supplement the master equations with covariance equations]
In order to find the scalar and tensor potentials satisfying the master equations, it is immensely useful to enforce the tensor-densital character of these objects by adding further linear homogeneous partial differential equations. As it will turn out, the appropriate partial differential equations contain terms that also appear in the master equations and may thus be used to great advantage. Most importantly these additonal equations will relieve us from having to worry about the tensor-densital character of the potentials when solving the master equations, since the enforcement of the corresponding transformation behaviour of the potentials under coordinate transformations will be taken care of precisely by these covariance equations.

The form of the covariance equations heavily depends on the index structure of the independent geometric tensor fields $G^A$, and hence must be derived on a case by case basis. Conceptually,  their derivation is straightforward. The key idea \cite{Rund} is to start from the required transformation behaviour of some particular hypersurface field and to derive it with respect to the highest (and then second highest, and so on, down to the zeroth) derivative of the Jacobian of an arbitary coordinate transformation, all to be evaluated at the identity transformation. The resulting linear homogeneous differential equations for the hypersurface field then encode the postulated transformation behaviour. This procedure is most transparently explained by way of a simple example, which is given in Appendix \ref{sec_covexample}.

The partial differential equations encoding the tensor-densital character of the scalar and tensor potentials are derived in precisely analogous fashion to the example given in the appendix, namely starting from the algebraic covariance equation for the scalar potential 
   $$C(T^A_{M}G^M,\partial(T^A_{M} G^M), \partial\partial(T^A_M G^M), \partial\partial\partial (T^A_M G^M)) = \det(T) C(G^A, \partial G^A, \partial\partial G^A, \partial\partial\partial G^A)\,,$$
where $T^A{}_M$ denotes the representation of the Jacobian as it acts on the geometric fields $G^M$, and the algebraic covariance equations for the tensor potentials
$$C_{B_1 \dots B_N}(T^A_{M}G^M,\partial(T^A_{M} G^M), \partial\partial(T^A_M G^M)) = \det(T) T^{C_1}_{B_1} \dots T^{C_N}_{B_N} C_{C_1 \dots C_N}(G^A, \partial G^A, \partial\partial G^A)\,,$$
by calculation of the derivatives of the above algebraic covariance equations for the potentials with respect to all appearing orders of derivatives of the Jacobian. 
There are four sets of covariance equations for the scalar potential $C$ (since this field depends on up to the third derivative of $G$) and three sets of covariance equations for the tensor potentials  $C_{B_1 \dots B_N}$ (since these all depend on at most the second derivative of $G$). The combined system of differential equations provided by these covariance equations together with the master equations then automatically selects all solutions that are tensor densities of weight one.

\item[Step 9. Solve the master and covariance equations]
The problem of finding gravitational dynamics for the coefficients of the matter equations we started from amounts to nothing more, but also nothing less, than finding solutions to the master equations combined with the covariance equations for the potentials $C$ and $C_{B_1 \dots B_N}$. Indeed, the physical question of whether there exist any gravitational dynamics at all which do not contradict the predictivity and quantizability of the specified matter equations reduces to the mathematical question of existence of solutions to the said linear homogeneous system of partial differential equations; likewise, the physical question of whether there are several such gravity theories reduces to the mathematical question of the uniqueness of solutions; finally the most interesting physical question, namely what the precise form of suitable gravitational dynamics are, reduces to the mathematical problem of finding a concrete solution.   

Our first case study shows that the master equations following from Maxwell electrodynamics feature as their unique solution the Einstein-Hilbert Lagrangian with undetermined gravitational and cosmological constants emerging as integration constants.  The second case study then shows that other (non-standard model) matter requires a different gravitational theory. 


\item[Step 10. Impose judicious choices of energy conditions and symmetry reductions]
Be\-yond the physically non-negotiable conditions that the matter equations be predictive and quantizable, one may impose further conditions on the matter dynamics, such as (strong, dominant, \dots) energy conditions on the Gotay-Marsden energy-momentum tensor density
$$T^{a}{}_b := J^{\mathcal{A}\, a}{}_b \frac{\delta S_\textrm{\small matter}}{\delta G^\mathcal{A}}\,,$$
where $\mathcal{A}$ stands for the indices carried by the spacetime geometric tensor $G$ and the intertwiners $J^{\mathcal{A} \, a}{}_b$ are read off the Lie derivative 
$$(\mathcal{L}_\xi G)^\mathcal{A} = J^{\mathcal{A}}{}_b \,\xi^b + J^{\mathcal{A}\, a}{}_b\, \xi^a{}_{,b}$$
for an arbitrary spacetime vector field $\xi$. While such additional conditions can have no bearing on the above form of the master equations (since the latter follow already from the predictivity and quantizability of the matter dynamics), they may serve to further restrict the geometric degrees of freedom, and thus reduce the equations correspondingly. 

Another strategy to simplify the master and covariance equations is to derive actions for spacetimes $(M,G)$ with Killing vector fields $K_1, \dots, K_n$, 
$$(\mathcal{L}_{K_i} G)^{\mathcal{A}} = 0 \qquad \textrm{ for } i=1,\dots ,n\,,$$
whose algebra $[K_i,K_j]=f^{k}{}_{ij} K_k$ gives rise to a negative definite Killing form
$$\mathcal{K}_{ij} := f^{m}{}_{ni} f^{n}{}_{mj}\,,$$
since in that case the corresponding symmetry group is compact, which suffices \cite{Palais} to ensures that the symmetry-reduced action yields the same equations of motion as would have been obtained by a symmetry-reduction of the field equations following from the full, not symmetry-reduced action.. 
Thus this strategy works for, e.g., spherical symmetry, but unfortunately not for homogeneous and isotropic spacetimes modelling simple cosmologies.  
\end{description}

With the above procedure to derive gravitational actions from specified matter actions in place, we turn to two concrete case studies in order to illustrate its application in vivo. 
\newpage
\section{First case study:\\ Gravity underlying maxwell theory}\label{sec:caseI}
The following application of the practical rules laid down in the previous section, to the case of Maxwell theory as the prescribed matter inhabiting the spacetime, serves as a warm-up exercise to the more ambitious case study presented in the next section. But since the result is the standard textbook Einstein-Hilbert action, with only the gravitational and cosmological constant left to be determined by experiment, this simplest possible case already illustrates the power of the master equations.  
\begin{description}
   \item[Step 1. Test matter] On a smooth four-dimensional manifold $M$, we consider matter described by a covector field $A$ obeying dynamics encoded in the Maxwell action
$$S_\textrm{\small Maxwell}[A,g] := -\frac{1}{4} \int d^4x \det(g)^{-1/2} g^{ac} g^{bd} F_{ab} F_{cd}\,,$$
where some non-degenerate symmetric $(2,0)$-tensor field $g$, employed to construct a scalar density from the field strength $F=dA$, provides an additional structure on $M$. Following the philosophy of this article, we make no further a priori assumptions about this tensor field $g$, neither technically nor concrning its physical role, since all physically required properties can be derived and thus must not be stipulated. According to the general parlance agreed upon in Step 1 of the general recipe, we refer to $g$ as the `geometry' on $M$, but without meaning anything more by this than that the geometry completely determines the coefficients of the matter field equations, as is manifest from the above action. 

\item[Step 2. Principal tensor field] The field equations for the covector field $A$ one derives from the above action features a gauge ambiguity that we choose to fix by imposing the gauge
$$\partial_a(\det(g)^{-1/2} g^{ab} A_b) = 0\,,$$
which yields the gauge-fixed equations of motion 
$$0=\det(g)^{1/2} g^{cd}\partial_{a_1} \left[\det(g)^{-1/2} g^{a_1 a_2} \partial_{a_2} A_{d}\right] = g^{cd} g^{a_1a_2}  \partial_{a_1} \partial_{a_2}  A_d + \textrm{ lower derivative terms}\,.$$
From the coefficient of the highest derivative term one reads off the principal tensor 
$$P(k) = \pm \omega \det_{c,d}\left[g^{cd} g^{a_1 a_2} k_{a_1} k_{a_2}\right] = \pm\omega \det(g) \left(g^{a_1 a_2} k_{a_1} k_{a_2}\right)^4\,,$$
which is de-densitized by letting $\omega = \det(g)^{-1}$, and upon removal of repeated factors simply becomes
$$P(k) = \pm g^{a_1 a_2} k_{a_1} k_{a_2}\,.$$

\item[Step 3. Dual tensor field] Since the principal tensor field is irreducible, we only need to consider one map
$$(DP(k))^a := \pm 2 g^{a m} k_{m}$$
and observe that  
$$P^\#(X) := g_{b_1 b_2} X^{b_1} X^{b_2}$$
satisfies the duality requirement for any covector $k$ with $P(k)=0$,
$$P^\#(DP(k)) = 4 g_{b_1 b_2} g^{b_1 m} g^{b_2 n} k_m k_n = 4 g^{mn} k_m k_n = 4 P(k) = 0\,.$$
Obviously, multiplying the above-defined dual tensor field with a real function on the manifold $M$ again provides a dual tensor field. This is of course a generic feature of the dual tensor, independent of the case presently studied, and all further constructions are independent of this ambiguity. 

\item[Step 4. Bi-hyperbolicity] The principal tensor field $P$ is easily shown to be hyperbolic and to satisfy the sign convention if and only if the $(2,0)$-tensor $g$ has Lorentzian signature $(+-\dots-)$. 

This can be seen as follows. If $P$ is hyperbolic, then there exists a hyperbolic $h$ with $P(h)>0$, so that the equation 
$$P(q+\lambda h)=\lambda^2 g^{ab} h_a h_b + 2 \lambda g^{ab} h_a q_b + g^{ab}q_a q_b = 0$$
has only real roots $\lambda$. But then the discriminant $(g^{ab} h_a q_b)^2 - g^{ab} h_a h_b g^{cd} q_c q_d$ of this equation is positive. Choosing a cotangent basis with $\epsilon^0:=h$ such that $g^{ab} \epsilon^0_a \epsilon^\alpha_b = 0$, one sees that $g^{ab}\epsilon^0_a \epsilon^0_b > 0$ and can further write the discriminant as $q_\alpha q_\beta g^{ab} \epsilon^\alpha_a \epsilon^\beta_b < 0$ for all $q_\alpha$, which proves that $g^{ab}$ has mainly minus Lorentzian signature.  Conversely, if $g$ is of the said signature, it is immediate that $P$ is hyperbolic, as one quickly sees in any $g$-orthonormal cotangent basis.  

Hyperbolicity of the dual tensor field is automatic in this case, since a metric has the same signature as its inverse.

\item[Step 5. Geometric degrees of freedom] We assume to be given a hypersurface in $M$ with an everywhere hyperbolic covector field $n$ normalized to $P(n)=1$ that annihilates any of three linearly independent tangent vector fields $e_1, e_2, e_3$, such that we construct complete spacetime tangent and co-tangent space bases
$$e^a_0:=g^{ab} n_b, \quad e_1,\quad e_2,\quad e_3 \qquad \textrm{ and } \qquad \epsilon^0 := n,\quad \epsilon^1,\quad \epsilon^2,\quad \epsilon^3$$
dual to each other, giving rise to independent geometric hypersurface tensor fields
$$g^{00} := g(\epsilon^0, \epsilon^0), \qquad g^{0\alpha} := g(\epsilon^0,\epsilon^\alpha),\qquad g^{\alpha\beta}=g(\epsilon^\alpha,\epsilon^\beta)$$
for $\alpha,\beta=1,2,3$, 
which by the normalization and annihilation properties are however constrained by
$g^{00} = 1$ and $g^{0\alpha} = 0$, 
so that we identify as the independent geometric degrees of freedom the symmetric non-degenerate hypersurface tensor field
$$G^A := (g^{\alpha\beta})\,.$$

\item[Step 6. Coefficients] According to the general rules, one calculates the coefficients
\begin{eqnarray*}
M^{\alpha_1 \alpha_2\, \gamma} &=& g^{0\alpha_2} P^{\alpha_1\gamma} + g^{\alpha_1 0} P^{\alpha_2\gamma} = 0\,,\\
U^{\alpha_1 \alpha_2 \, \rho \chi} &=& -P^{\chi\alpha_1} g^{\rho\alpha_2} - P^{\chi\alpha_2} g^{\alpha_1\rho} = - 2 g^{\chi(\alpha_1} g^{\alpha_2)\rho} \,,\\
V^{\alpha_1\alpha_2\, \chi} &=& P^{\chi\lambda} \partial_\lambda g^{\alpha_1\alpha_2} + P^{\chi\alpha_1} \partial_\lambda g^{\lambda \alpha_2} + P^{\chi\alpha_2} \partial_\lambda g^{\alpha_1\lambda} = g^{\chi\lambda} \partial_\lambda g^{\alpha_1\alpha_2} +  2 g^{\chi(\alpha_1} \partial_\lambda g^{\alpha_2)\lambda}
\end{eqnarray*}
and thus obtains the further coefficients
\begin{eqnarray*}
  Q_{\alpha_1\alpha_2}{}^{\beta_1\beta_2 \, \gamma} &=& 0\,,\\
  T^{\alpha_1\alpha_2 \, [\mu\nu]} &=& - 2 g^{[\nu|(\alpha_1} g^{\alpha_2)|\mu]} = 0\,,\\
  S^{\alpha_1\alpha_2\,\gamma} &=& - g^{\gamma \lambda} \partial_\lambda g^{\alpha_1\alpha_2} + 2 \partial_\lambda g^{\gamma(\alpha_1}  g^{\alpha_2)\lambda}\,. 
\end{eqnarray*}
   
\item[Step 7. Master equations]
With the coefficients calculated above, the first master equation takes the form
$$\frac{\partial C}{\partial \partial_{\beta_1\beta_2\beta_3} g^{\alpha_1\alpha_2}} = 0\,,$$
so that we immediately learn that, in the present case, even the scalar potential $C$ depends only on $g$, $\partial g$ and $\partial\partial g$, but not the third derivative $\partial\partial\partial g$. Further, the master equations ($5$) and ($6$) are identically satisfied, and so are the two sequences of master equations ($7_N$) and ($8_N$) for all $N\geq 1$. The remaining equations are\\[6pt]
\begin{tabular}{ll}
 (4') & $0 = 2 \partial_\mu(C_A U^{A\,(\beta\mu)}) +2 C_A S^{A\,\beta}\,,$\\
    ($9_N'$) & $0 = \frac{\partial C_{B_1 \dots \widetilde{B_i} \dots B_N}}{\partial \partial^2_{\mu\nu}G^{B_i}} - \frac{\partial C_{B_1 \dots \dots B_{N-1}}}{\partial \partial^2_{\mu\nu}G^{B_N}} \qquad \textrm{with the index } \widetilde{B_i} \textrm{ removed for } i=1,\dots,N \,,$\\     
($10_N'$)  &  $0 =  (N+1)!\, C_{AB_1\dots B_N} U^{A\,(\alpha\beta) }  -(N-2)(N-1)!\, \frac{\partial C_{B_1 \dots B_{N-1}}}{\partial \partial^2_{\alpha\beta} G^{B_N}}\,, $\\
    ($11_N'$) \quad\quad & $0 = (N+1)! C_{A B_1 \dots B_N}  S^{A\,\alpha}+ (N-1)! \sum_{a=1}^N \frac{\partial C_{B_1 \dots \widetilde{B_a} \dots B_N}}{\partial\partial_{\alpha} \hat{G}^{B_a}} 
           - 2(N-1)! \partial_\gamma  \frac{\partial C_{B_1 \dots B_{N-1}}}{\partial\partial^2_{\alpha\gamma} \hat{G}^{B_N}}\,,$
\end{tabular}

for $N\geq 1$. Note that the master equations ($2$) and ($3$) are contained in the last two sequences as the special case $N=1$.

\item[Step 8. Covariance equations] Since in the present case both the scalar and the tensor potentials depend on at most second derivatives of the geometric hypersurface tensor field, the covariance equations take the same form for all $N\geq 0$, namely \\[6pt]
\begin{tabular}{ll}
(\textrm{Cov}2) & $0=g^{\alpha (\sigma}\,\frac{\partial C_{B_1\dots B_N}}{\partial\partial^2_{\mu\nu)}  g^{\alpha\rho}}$\,,\\
(\textrm{Cov}1) & $0=2\, g^{\alpha (\mu}\frac{\partial C_{B_1\dots B_N}}{\partial \partial_{\nu)}  g^{\alpha\rho}}-\partial_\rho  g^{\alpha\beta}\frac{\partial C_{B_1\dots B_N}}{\partial\partial^2_{\mu\nu} g^{\alpha\beta}}+\,4\,\partial_{\sigma} g^{\alpha(\mu}\frac{\partial C_{B_1\dots B_N}}{\partial\partial^2_{\nu)\sigma}g^{\alpha\rho}}\,,$
\end{tabular}

which are obtained from deriving the algebraic transformation law for the weight-one tensor densities $C_{B_1 \dots B_N}$ for $N\geq 0$ with respect to the second and first derivatives of the Jacobian of a coordinate transformation. The third covariance equation $(\textrm{Cov}0)$ is not displayed since in the present case it is not required for a solution of the master equations.  

\item[Step 9. Solution of the master and covariance equations]
Now we can solve the master equations step by step. First, we observe that equation ($10_N'$) for $N=2$ simply reads
\begin{equation*}
0=C_{\rho\sigma\alpha_1\beta_1\alpha_2\beta_2} U^{\rho\sigma\,\mu\nu}\,,
\end{equation*}
which may be solved to yield $C_{\rho\sigma\alpha_1\beta_1\alpha_2\beta_2}=0$. Inserting this result back into equation ($10_N'$), first for $N=4$ and then repeating the procedure for all even $N$, we see that all potentials with an odd number of index pairs already vanish, except for the first one, $C_{\alpha\beta}$. For our next conclusion, we temporarily change variables in favour of the metric $ g_{\alpha\beta}$. Changing the partial deriviatives of $ g^{\alpha \beta}$ accordingly, the covariance equation $(\textrm{Cov}2)$ becomes
\begin{equation*}
0=\frac{\partial C_{\alpha_1\beta_1\dots\alpha_N\beta_N}}{\partial  g_{\alpha(\beta,\gamma\delta)}}\,,
\end{equation*}
where we denote partial derivatives by a comma. Moreover, the divergence term in equation ($11_N'$) implies
\begin{equation*}
0=\frac{\partial^2 C_{\alpha_1\beta_1\dots\alpha_N\beta_N}}{\partial  g_{\alpha\beta,(\mu\nu|}\,\partial  g_{\rho\sigma,|\gamma)\delta}}\,.
\end{equation*}
But the last two equations already yield 
\begin{equation*}
0=\frac{\partial^2 C_{\alpha_1\beta_1\dots\alpha_N\beta_N}}{\partial  g_{\alpha\beta,\mu\nu}\,\partial  g_{\rho\sigma,\gamma\delta}}\qquad \textrm{ (without symmetrization)}\,,
\end{equation*}
implying that all remaining potentials $C_{\alpha_1\beta_1\dots\alpha_N\beta_N}$  can only depend at most linearly on the second derivatives of the field $ g_{\alpha\beta}$ and similarly of $ g^{\alpha\beta}$. Since, in particular, the scalar potential $C$ depends only linearly on the second derivatives of $ g^{\alpha\beta}$, we conclude from equation ($10_N'$) for $N=1$ that the potential $C_{\alpha_1\beta_1\alpha_2\beta_2}$  must in fact be independent of the second derivatives of $ g^{\alpha\beta}$. Using this result in equation ($10_N'$) for $N=3$, and iterating on all odd $N$, we find that also all even potentials $C_{\alpha_1\beta_1\dots\alpha_N\beta_N}$ for $N\geq4$ vanish. Hence, it only remains to determine the potentials $C$, $C_{\alpha\beta}$ and $C_{\alpha\beta\gamma\delta}$.

As described in Appendix \ref{sec:Invar}, we may now perform a change from the arguments $(g^{\alpha\beta},\partial_\mu g^{\alpha\beta}, \partial_\mu\partial_\nu g^{\alpha\beta})$, on which the tensor and scalar potentials depend, to a set of arguments $(g^{\alpha\beta}, R_{\alpha\beta\gamma\delta})$, where $R_{\alpha\beta\gamma\delta}$ is the Riemann-Christoffel tensor of $ g^{\alpha\beta}$, such that the covariance equations are automatically solved if and only if $C_{B_1 \dots B_N} = C_{B_1\dots B_N}(g_{\alpha\beta}, R_{\alpha\beta\gamma\delta})$ for all $N\geq 0$. In three dimensions, we know that the Riemann tensor can be expressed in terms of the Ricci tensor $R_{\alpha\beta}$ and the metric $ g_{\alpha\beta}$ so that, actually, $C_{B_1\dots B_N}=C_{B_1\dots B_N}( g^{\alpha\beta}, R_{\alpha\beta})$. The only such scalar density of weight one that is linear in the Ricci tensor (recall that the at most linear dependence of the potentials on the Riemann tensor did not follow from the covariance equations alone, but involved one of the master equations) is $(-\det  g)^{-1/2} R$, with the Ricci scalar $R=R_{\alpha\beta}  g^{\alpha\beta}$, and the minus sign under the square root accounts for the fact that $ g^{\alpha\beta}$ must be negative definite. Thus we arrive at
\begin{equation*}
C=-(2 \kappa)^{-1} (-\det  g)^{-1/2}\,(R-2\lambda),
\end{equation*}
with constants $\kappa$ and $\lambda$, as the only scalar potential that meets all the requirements. 

Then we can immediately calculate, from equation ($10_N'$) for $N=1$, that 
\begin{equation*}
C_{\alpha\beta \mu\nu} =  (16\kappa)^{-1}(-\det  g)^{-1/2} \left[
 g_{\alpha \mu} g_{\beta \nu} + g_{\beta \mu} g_{\alpha \nu} - 2  g_{\alpha\beta} g_{\mu\nu}\right]\,.
\end{equation*}
In terms of the $( g^{\alpha\beta}, \Gamma^{\alpha}_{\beta\gamma}, R_{\alpha\beta\gamma\delta})$, the coefficient $S^{\alpha\beta\,\gamma}$ can be rewritten as
\begin{equation*}
S^{\alpha\beta\,\gamma}=U^{\alpha\beta\,\mu\nu}\Gamma^{\gamma}_{\mu\nu}\,,
\end{equation*}
which makes it easy to see that equation ($4'$) takes the form
\begin{equation*}
0= g^{\mu\rho} g^{\sigma\nu}\nabla_\nu C_{\rho\sigma}\,,
\end{equation*}
where $\nabla_\gamma$ denotes the covariant derivative with respect to the Levi-Civita connection. Using the well-known theorem due to Lovelock \cite{Lovelock:1972vz}, which also for the case of three dimensions asserts that the only divergence-free second rank tensor depending only on the metric and its first and second derivatives is the Einstein tensor, and the fact that again $C_{\rho\sigma}$ can only depend linearly on the Ricci tensor, we immediately conclude that
\begin{equation*}
C_{\alpha\beta}=\beta_1(-\det  g)^{-1/2}\,(R_{\alpha\beta}-\frac{1}{2}\, g_{\alpha\beta}\, R)+\beta_2(-\det  g)^{-1/2}\, g_{\alpha\beta}\,.
\end{equation*}
The remaining master equations ($9'$) and ($11_N'$) are then identically satisfied. 

The potentials $C$, $C_{\alpha\beta}$ and $C_{\alpha\beta\gamma\delta}$ derived above completely determine the Lagrangian by virtue of
\begin{equation*}
L=C_{\alpha\beta\gamma\delta} K^{\alpha\beta} K^{\gamma\delta}+C_{\alpha\beta}  K^{\alpha\beta} + C
\end{equation*}
and thus we have found the gravitational dynamics of the geometry $ g^{\alpha\beta}$. However, one may simplify this result a little further. 
We immediately realize that the potential $C_{\rho \sigma}$ can be written as the functional derivative of the scalar density
\begin{equation*}
\Lambda=\beta_1(-\det  g)^{-1/2}\,R-2\,\beta_2\,(-\det  g)^{-1/2}
\end{equation*}
with respect to $g^{\alpha\beta}$. This has severe consequences for the relevance of this potential in the equations of motion displayed in the general description of Step 7. The part of the Lagrangian involving $\Lambda$ satisfies the equations of motion identically and is thus dynamically irrelevant \cite{Kuchar:1974es}. This can be seen as follows. The kinematical relation supplementing the Lagrangian equations of course remains untouched because it is independent of the Lagrangian, so we have that 
\begin{equation*}
\dot{ g}^{\alpha\beta}(z)=N(z)  K^{\alpha\beta}(z) + (\mathcal L_{\vec N}  g)^{\alpha\beta}(z)\,.
\end{equation*}
The actual Lagrangian equation of motion reads
$$\frac{\partial}{\partial t} \left(\frac{\partial L(z)}{\partial  K^{\alpha\beta}(z)}\right)=\int_\Sigma\, dx\,\left[N(x)\frac{\delta L(x)}{\delta  g^{\alpha\beta}(z)}\right]+\mathcal L_{\vec N}\left(\frac{\partial L(z)}{\partial  K^{\alpha\beta}(z)}\right)
$$
in this case,  because there is no contribution from the coefficients $Q_A{}^{B\,\gamma}$. We may now insert the part $L_{\textrm{lin}}(z):=\delta \Lambda(z)/\delta \hat P^{\alpha\beta}(z) \hat K^{\alpha\beta}(z)$ of the Lagrangian that is  linear in the velocities $ K^{\alpha\beta}$ into the left hand side of this equation in order to find, taking into account the kinematical supplement, that
$$L_{\textrm{lin}}(z)= \int_\Sigma\, dx\, \frac{\delta^2\Lambda(z)}{\delta  g^{\rho\sigma}(x)\delta  g^{\alpha\beta}(z)}\,\left(N(x)  K^{\rho\sigma}(x) + (\mathcal L_{\vec N}  g)^{\rho\sigma}(x)\right)\,.$$
It is then straightforward to see that these are terms of precisely the form as those appearing on the right hand side of the previous equation. The respective first terms cancel because the functional derivatives commute. That also the second terms cancel, one can see by writing out the Lie derivative on both sides and using the chain rule and an integration by parts on the left hand side of the equation. 


It is instructive to convert the thus obtained Lagrangian to a Hamiltonian, to which end we calculate the canonical momenta as the Legendre dual variables of the velocities,
\begin{equation*}
 \pi_{\alpha\beta}=\frac{\partial L}{\partial  K^{\alpha\beta}}=2 C_{\alpha\beta\gamma\delta} K^{\gamma\delta}+\frac{\delta \Lambda}{\delta  g^{\alpha\beta}}\,,
\end{equation*}
where we again included the term $\Lambda$ discarded above, just in order to see how that it can be discarded in the canonical picture equally well, since the Poisson brackets on the geometric phase space spanned by $(g^{\alpha\beta}, \pi_{\alpha\beta})$ do not change if we add to the canonical momenta the functional derivative of a weight-one scalar density with respect to the configuration variables $G^A$. Thus, we can redefine the canonical momenta,
\begin{equation*}
 \pi_{\alpha\beta}\rightarrow \tilde{\pi}_{\alpha\beta}= \pi_{\alpha\beta}-\frac{\delta \Lambda}{\delta  g^{\alpha\beta}}\,
\end{equation*}
and invert the second last equation to get the velocities
\begin{equation*}
 K^{\alpha\beta}=\frac{1}{2} C^{\alpha\beta\gamma\delta} \tilde{\pi}_{\alpha\beta}\,,
\end{equation*}
where $C^{\alpha\beta\gamma\delta}$ is the inverse of the potential $C_{\alpha\beta\gamma\delta}$ and explicitly reads
\begin{equation*}
C^{\alpha\beta\gamma\delta}=4 \kappa \,(-\det  g)^{1/2}\,( g^{\alpha\gamma} g^{\beta\delta}+ g^{\beta\gamma} g^{\alpha\delta}- g^{\alpha\beta} g^{\gamma\delta})\,,
\end{equation*}
which is known as the DeWitt tensor density. The local superhamiltonian then automatically becomes
\begin{eqnarray*}
{\mathcal H}_{\textrm{local}}&=& K^{\alpha\beta}\tilde{\pi}_{\alpha\beta} - C_{\alpha\beta\gamma\delta} K^{\alpha\beta} K^{\gamma\delta}-C\nonumber\\
&=&\frac{1}{4}C^{\alpha\beta\gamma\delta}\tilde\pi_{\alpha\beta}\tilde\pi_{\gamma\delta}
+(2\kappa)^{-1}(-\det  g)^{-1/2}\,(R-2\lambda)\,,
\end{eqnarray*}
which is the famous Arnowitt-Deser-Misner Hamiltonian \cite{arnowitt1960canonical} of Einstein-Hilbert dynamics with a cosmological term,
$$S_\textrm{\small grav}[g] = \frac{1}{2\kappa}\int_M d^4x \sqrt{-\det g}\, ( R-2\lambda)\,,$$
where $g$ is the spacetime metric and $R$ the associated spacetime Ricci scalar.  

\item[Step 10. Additional energy or symmetry conditions] were not needed to obtain an analytic solution of the master equations in this case. 
\end{description}

In summary, we arrived at the interesting conclusion that the unique gravitational dynamics for a  four-dimensional metric spacetime $(M,g)$ carrying predictive and quantizable Maxwell electrodynamics is given by the familiar Einstein-Hilbert dynamics for a Lorentzian metric, with undetermined gravitational and cosmological constants appearing as integration constants when solving the master equations. This result directly extends to matter dynamics $S_{\textrm{SM}}[g,\Phi]$ including all fields of the standard model of particle physics, because their equations of motion all share the same principal tensor fields, which by deliberate construction of the standard model (taking particles to be the irreducible representations of the local Lorentz group) is precisely the principal tensor field of Maxwell electrodynamics.

\newpage
\section{Second case study:\\ Gravity underlying $SO(p,q)$-violating fermionic matter}\label{sec:diracexample}\label{sec:caseII}
In order to see the machinery to derive gravitational dynamics underpinning particular matter dynamics working at full capacity, we will now consider a vector-tensor geometry $(M,g,W)$, constituted by a metric and a vector field $W$, and find gravitational dynamics for it such that an $SO(p,q)$-violating extension of Dirac dynamics is predictive and quantizable on that geometry. While we are of course {\it not} proposing either this particular geometry nor this particular type of matter equations as a model for any observable physics, but rather as a deliberately brutal---but nevertheless causally fully consistent---deviation from standard model physics, this case well illustrates that even such matter dynamics can be underpinned by suitable gravitational dynamics. 

\begin{description}
\item[Step 1. Test matter] As test matter dynamics we now directly stipulate $SO(p,q)$-violating  field equations 
\begin{equation*}
(i \gamma^a + W^a) D_a \Psi = 0\,
\end{equation*}
for a spinor field $\Psi$ on a four-dimensional smooth manifold equipped with a geometry $(g,W)$ consisting of a spacetime metric $g$ (of a so far arbitrary but fixed signature $(p,q)$, which will be considerably restricted by the bi-hyperbolicity condition in Step 4) together with a spacetime vector field $W$. The spacetime $\gamma$-matrices $\gamma^a=\gamma^I E^a_I$ are constructed with the help of local frame fields $E_I$ satisfying $g^{ab}=\eta^{IJ}E^a_I E^b_J$ and the flat spacetime $\gamma$-matrices $\gamma^I$ satisfying the Clifford algebra $\{\gamma^I,\gamma^J\}=2 \eta^{IJ}$, where $\eta^{IJ}=\textrm{diag}(1, \dots, 1, -1, \dots, -1)^{IJ}$ with the same signature as $g$. We assume that the spacetime admits a spin structure (whose existence is of course still equivalent to the vanishing of the second Stiefel-Whitney class associated with the $g$-orthonormal frame bundle over $M$) such that the spin covariant derivative $D_a$ is induced from the torsion-free spin connection by virtue of
\begin{equation*}
^{S}\Gamma^I_{a J}= - E^{b}_{J} (\partial_{a} \theta^{I}_{b} - \Gamma^{c}_{ab}\,, \theta^{I}_c )
\end{equation*}
where $\Gamma^{c}_{ab}$ are the Christoffel symbols of the metric $g^{ab}$, and $\theta^{I}_{b}$ denote the coframe fields dual to the frame fields $E^a_I$. The spin connection is antisymmetric with respect to $\eta^{IJ}$, and 
\begin{equation*}
D_a=\partial_a-\frac{i}{4} {}^{S}\Gamma^I_{a J} \,\eta_{IK} \,[\gamma^{K},\gamma^J]
\end{equation*}
if the covariant derivative acts on spinors $\Psi$. Here and in the following, we will suppress all spinor indices.

\item[Step 2. Principal tensor field] By acting on the equations of motion with the differential operator $(i \gamma^J E^{b}_{J} - W^b) D_b$ from the left, we obtain the equation
\begin{equation*}
-(\gamma^J \gamma^{I} E^{b}_J E^{a}_I + W^a W^b - i \gamma^J E^b_J W^a + i \gamma^I E^a_{I} W^b) D_b D_a \Psi +
 i \gamma^{J} E^b_J D_b W^a D_a \Psi=0\,,
\end{equation*}
from whose highest order derivative terms we obtain, using the Clifford algebra relation $\{\gamma^I,\gamma^J\}=2 \eta^{IJ}$ and the fact that partial derivatives commute, the principal tensor field
\begin{equation*}
P^{a b} = (g^{a b} + W^a W^b)\,.
\end{equation*}

\item[Step 3. Dual tensor field] 
Since the principal tensor has rank two, it is again simple to calculate a dual tensor field $P^\#(x,v)$ in terms of the inverse of the matrix $g^{ab}+W^aW^b$. Indeed, one quickly finds the dual tensor  
$$P^\# (x,v) =\left (g_{a b} - \frac{1}{1+ W^r W^s g_{rs}} W^m W^n g_{m a} g_{n b}\right )  v^a v^b$$
with respect to the principal tensor $P$. 

\item[Step 4. Bi-hyperbolicity] 
The hyperbolicity and signature condition on the principal tensor now simply amount to the algebraic requirement that the matrix $g^{ab}+W^aW^b$ have mainly minus Lorentzian signature at every point of the manifold. However, this does of course by no means imply that the metric $g$ itself has to be of Lorentzian signature. In fact, the principal tensor is hyperbolic in two different cases: either the metric $g$ has signature $(+\,-\,-\,-)$ and the vector field $W$ is timelike, or null, or of spacelike length $-g(W,W)<1$ with respect to $g$, or the metric has signature $(-\,-\,-\,-)$ and the vector field has length $-g(W,W)>1$. Interestingly, the two cases differ in the way hyperbolicity is encoded in the geometry. In the first case, hyperbolicity is ensured by the metric, whereas in the second case, it is the vector field which renders the combination $g^{ab}+W^a W^b$ hyperbolic.

The hyperbolicity of the dual polynomial is in this case again equivalent to the hyperbolicity of the principal polynomial so that, also here, bi-hyperbolicity does not enforce further algebraic constraints on the values of $g$ and $W$ beyond what is already enforced by hyperbolicity. 


\item[Step 5. Geometric degrees of freedom] We assume to be given a hypersurface in $M$ with an everywhere hyperbolic covector field $n$ normalized to $P(n)=1$ that annihilates any of three linearly independent tangent vector fields $e_1, e_2, e_3$, such that we construct complete spacetime tangent and co-tangent space bases
$$e^a_0:=(g^{ab}+W^aW^b) n_b, \quad e_1,\quad e_2,\quad e_3 \qquad \textrm{ and } \qquad \epsilon^0 := n,\quad \epsilon^1,\quad \epsilon^2,\quad \epsilon^3$$
dual to each other, giving rise to independent geometric hypersurface tensor fields
$$ g^{\alpha \beta}:= g(\epsilon^\alpha,\epsilon^\beta)\,, \quad g^{0\alpha}:= g( \epsilon^0,\epsilon^\alpha)\,, \quad g:=g(\epsilon^0,\epsilon^0)\,,\quad
W^\alpha:= \epsilon^\alpha(W)\,, \quad W^0:=  \epsilon^0(W)\,.$$
However, not all of these hypersurface tensors can be independent since the frame conditions $P(n)=1$ and $e_0(\epsilon^\alpha)=0$ can be used to express $W$ and $W^\alpha$ in terms of the projections $g$ and $g^{\alpha}$. Thus, the hypersurface tensor fields $g$, $g^\alpha$ and $g^{\alpha \beta}$ already constitute a possible parametrization of the spacetime geometry $(g,W)$. Indeed, one can check that the completeness relations
\begin{eqnarray*}
g^{a b}&=& g\, e_0^a e_0^b + 2 g^{\alpha}\, e_0^{(a}\,e^{b)}_{\alpha}+ g^{\alpha \beta}\, e^a_{\alpha} e^{b}_{\beta} \quad \text{and}\\
W^a&=& \pm (1-g)^{1/2} \,e_0^a \mp \frac{1}{(1-g)^{1/2}}\, g^\alpha\, e^a_{\alpha}
\end{eqnarray*}
allow for a reconstruction of the spacetime geometry on the hypersurface, and in particular of the hypersurface tensor field 
$$P^{\alpha\beta} = g^{\alpha\beta} + \frac{1}{1-g}g^\alpha g^\beta \,.$$

 In principle, one could now choose $G^A:=(g^{\alpha\beta},g^\alpha,g)$ as the independent degrees of freedom and press on to the next step and determine the coefficients for the master equations. In particular, one would obtain the coefficients
\begin{eqnarray*}
 M^{\alpha\beta\,\gamma} &=& 2 g^{(\alpha} g^{\beta) \gamma} + \frac{2}{1-g} g^\alpha g^\beta g^\gamma\\
 M^{0\alpha\, \gamma} &=& \frac{g}{1-g}\, g^\alpha g^\gamma - (1-g) g^{\alpha \gamma}\\
 M^{00\,\gamma} &=& - 2 g^\gamma\,,
\end{eqnarray*}
which produce correct, but unnecessarily complicated master equations.
A more advantageous choice of configuration variables (as we will see when calculating the associated coefficients in the next step) is obtained by the field redefinitions
\begin{equation*}
\begin{array}{ccc}

\left\{\begin{array}{rcl}
P^{\alpha \beta} & := & g^{\alpha \beta} + \frac{1}{1-g} g^\alpha g^\beta\,, \\
g_{\alpha} & := & - \frac{1}{1-g} P_{\alpha \gamma} \,g^{\gamma}\,,\\
\phi & := & 1-g+ \frac{g^\alpha g^\beta g_{\alpha \beta}}{1-g-g^\alpha g^\beta g_{\alpha \beta}}
\end{array} \right\}
& \textrm{recovering} &
\left\{\begin{array}{rcl}
g^{\alpha \beta}&=& P^{\alpha \beta} - \frac{\phi}{1+ P^{\rho \sigma} g_{\rho} g_{\sigma}} P^{\alpha \gamma} g_{\gamma} P^{\beta \delta} g_{\delta}\,,\\
g^{\alpha}&=& -\frac{\phi}{1+P^{\gamma \delta} g_\gamma g_\delta} P^{\alpha \rho} g_{\rho}\,,\\
g &=& 1- \frac{1}{2} \phi - \sqrt{\frac{\phi^2}{4} - \phi^2 \frac{P^{\alpha \beta} g_{\alpha} g_\beta} {1+ P^{\gamma \delta} g_{\gamma} g_{\delta}}}\,.
\end{array}\right\}
\end{array}
\end{equation*}
We thus choose as the unscontrained geometric hypersurface tensor fields
$$G^A_{\textrm{\tiny redef}} := (P^{\alpha}, g_{\alpha}, \phi)\,.$$
 
\item[Step 6. Coefficients] 
The coefficients associated with the redefined fields $G_{\textrm{\tiny redef}}^A$, which are now functions of the original hypersurface fields $G^A$ obtained by projection from spacetime tensors, must be calculated from the coefficients of the projected fields according to product and chain rules, as explained in the general rules. In particular, we obtain
\begin{eqnarray*}
M_{\textrm{\tiny redef}}{}^{\alpha\beta\, \gamma} &=& M^{\alpha\beta\,\gamma} - \frac{(-1)}{(1-g)^2} M^{00\,\gamma} g^\alpha g^\beta + \frac{2}{1-g} g^{(\alpha} M^{|0|\beta)\, \gamma}=\dots=0\\
M_{\textrm{\tiny redef }\alpha}{}^\gamma &=& \frac{1}{(1-g)^2} M^{00\,\gamma} P_{\alpha\delta} g^\delta + \frac{1}{1-g} P_{\alpha\sigma} P_{\delta \rho} M_{\textrm{\tiny redef}}^{\sigma\rho\,\gamma} g^\delta -\frac{1}{1-g} P_{\alpha\delta} M^{0\delta\,\gamma} =\cdots = P^{\gamma\mu} g_\mu g_\alpha + \delta^\gamma_\alpha\,,\\[6pt]
M_{\textrm{\tiny redef}}{}^{\gamma} &=& \cdots = 0\,.
\end{eqnarray*}
The vanishing of the first and third set of coefficients and the simple form of the second set were the rationale behind the field redefinition made in the previous step.  Also according to the product and chain rule, one determines the coefficients
\begin{eqnarray*}
U_{\textrm{\tiny redef}}{}^{\alpha\beta\, \rho\chi} &=& U^{\alpha\beta\,\rho\chi} + \frac{1}{(1-g)^2} U^{00\,\rho\chi} + \frac{2}{1-g} g^{(\alpha} U^{|0|\beta)\,\rho\chi} = -2 P^{\chi(\alpha} P^{\beta)\rho}\,,\\
U_{\textrm{\tiny redef}\, \alpha}{}^{\rho\chi} &=& \frac{1}{(1-g)^2} U^{00\,\rho\chi} P_{\alpha\delta} g^\delta + \frac{1}{1-g} P_{\alpha\mu} P_{\delta\nu} U^{\mu\nu\,\rho\chi} g^\delta - \frac{1}{1-g} P_{\alpha\delta} U^{0\delta\,\rho\chi} = \delta^{\rho}_\alpha P^{\chi \mu} g_\mu\,,\\[6pt]
U_{\textrm{\tiny redef}}{}^{\rho\chi} &=& \cdots = 0
\end{eqnarray*}
and the coefficients
\begin{eqnarray*}
V_{\textrm{\tiny redef}}{}^{\alpha\beta\,\chi} &=& P^{\chi\gamma} P^{\alpha\beta}{}_{,\gamma}+ 2 P^{\chi(\alpha}\hat P^{\beta)\gamma}{}_{,\gamma}\,,\\
 V_{\textrm{\tiny redef } \alpha}{}^{\chi} &=& P^{\gamma\chi} g_{\alpha,\gamma} - P^{\chi\gamma}  g_{\gamma,\alpha}\,,\\
V_{\textrm{\tiny redef}}{}^{\chi} &=& P^{\chi\gamma} \phi_{\,,\gamma}\,.
\end{eqnarray*}
From the above nine sets of coefficients one then obtains directly, as the only non-vanishing coefficients $Q$,
\begin{eqnarray*}
Q_{\textrm{\tiny redef }\rho\sigma\, \alpha}{}^\gamma &=& -\frac{\partial M_{\textrm{\tiny redef }\alpha}{}^\gamma}{\partial P^{\rho\sigma}} = - \delta^\gamma_{(\rho} g_{\sigma)} g_\alpha\,\\
Q_{\textrm{\tiny redef}}{}^{\rho}{}_{\alpha}{}^{\gamma} &=&  -\frac{\partial M_{\textrm{\tiny redef }\alpha}{}^\gamma}{\partial g_\rho} = -P^{\gamma\rho} g_\alpha - P^{\gamma\mu} g_\mu \delta^\rho_\alpha\,,
\end{eqnarray*}
only vanishing coefficients $T_{\textrm{\tiny redef }}{}^{A\, [\mu\nu]} = 0$, and finally the coefficients
\begin{eqnarray*}
 S_{\textrm{\tiny redef }}{}^{\alpha \beta\,\mu}&=&- P^{\mu\gamma}  P^{\alpha \beta}{}_{,\gamma} + 2  P^{\gamma (\alpha}  P^{\beta)\mu}{}_{,\gamma}\,,\\
S_{\textrm{\tiny redef }}{}^{0\,\mu}&=&-  P^{\mu \gamma}  \phi_{,\gamma}\,,\\
S_{\textrm{\tiny redef }\alpha}{}^{\mu}&=& - 2 \, \delta^{(\mu}_{\alpha}  P^{\nu\sigma)}{}_{,\nu} g_\sigma - 2 \, \delta^{(\mu}_{\alpha}  P^{\nu\sigma)} g_{\sigma,\nu}  -  g_\alpha  g^\mu  ( P^{\nu \beta}  g_{\beta,\nu} + 2  g_\beta  P^{\nu\beta}{}_{,\nu}) \\
& & -  g_{\alpha, \nu} ( P^{\mu\sigma} P^{\nu\tau} g_\sigma  g_\tau + 2  P^{\mu \nu}) +  P^{\mu \nu}  g_{\nu,\alpha} -  g_{\alpha}  P^{\mu\nu}{}_{,\nu}\,.
\end{eqnarray*}

\item[Step 7. Master equations] Insertion of the coefficients calculated in the previous step into the master equations almost immediately leads to a number of drastic simplifications, which we will derive in the following three paragraphs, before writing down, in a fourth paragraph, the resulting reduced master equations that remain in the present case. 

\paragraph{Vanishing of the potentials $C^\mu$.} The master equations defined by the coefficients calculated above imply that the scalar potential $C$ only depends on at most the second partial derivatives of the geometric degrees of freedom. The simplest way to see this is
to trade in the first and second partial derivatives of the fields $P^{\alpha\beta}$, $\phi$ and $g_{\alpha}$ (which appear in the tensor potentials $C_{B_1 \dots B_{N\geq1}}$) for covariant derivatives with respect to the Levi-Civita connection $\Gamma^{\alpha}_{\beta\gamma}$ of the inverse metric $ P^{\alpha \beta}$ as well as the Riemann tensor $R_{\mu \nu \rho \sigma}$ and the non-tensorial quantity $S_{\mu\nu \rho \sigma}$ introduced in appendix \ref{sec:Invar}. The new fields  $\Gamma^{\alpha}_{\beta\gamma}$, $R_{\alpha\beta\gamma\delta}$ and $S_{\alpha\beta\gamma\delta}$ are then given in terms of $P^{\alpha\beta}$ and its partial derivatives by equations (\ref{Var1})-(\ref{Var3}) and the corresponding inverse transformations by (\ref{inverseVar1}) and (\ref{inverseVar2}), while the symmetrized first and second covariant derivatives of the fields $\phi$ and $g_\alpha$ are given in terms of the respective partial derivatives by equations (\ref{Var4})-(\ref{Var6}) and the inverse transformations by (\ref{inverseVar4})-(\ref{inverseVar6}).

In order to see that the potential $C$ in the present case does not depend on the third partial derivatives of the fields $G^A$, it is sufficient to rewrite the third partial derivatives of only the field $ g_\alpha$ in covariant form. The corresponding transformation formula is given by
\begin{eqnarray}
 g_{\alpha;(\beta\gamma\delta)}&=& g_{\alpha,\beta\gamma\delta}+  g_{\mu,\nu\lambda} (-3 \delta^\lambda_{(\delta} \delta^{\nu}_{\gamma} \Gamma^{\mu}_{\beta)\alpha}-3 \delta^\mu_\alpha \delta_{(\beta}^{\nu} \Gamma^{\lambda}_{\gamma\delta)})+ \textrm{lower order terms}\,,\nonumber
\end{eqnarray}
where, as we will see, it will not be necessary for our calculation to write out all terms of lower derivative order in $ g_\alpha$. 
The third partial derivatives of $g_\alpha$ can be recovered from the previous expression by employing the
\begin{equation*}
\textrm{useful identity} \qquad\Gamma^{\alpha}_{\beta\mu,\nu}=\Gamma^{\alpha}_{(\beta\mu,\nu)}-\frac{2}{3} R^{\alpha}{}_{(\beta\mu)\nu} + \frac{2}{3} \Gamma^{\alpha}_{\rho(\beta}\Gamma^{\rho}_{\mu)\nu} -\frac{2}{3} \Gamma^{\alpha}_{\nu\rho} \Gamma^{\rho}_{\beta\mu}\,.
\end{equation*} 

We can now rewrite the master equations in covariant form. We begin with the master equations (4)-(6) containing the potential $C$ and the potential $C_{A}$. Master equation (6) can be straightforwardly rewritten covariantly, but the chain rule in the first term in conjunction with the above expression for the symmetrized third covariant derivatives of $g_\alpha$, the derivative of the coefficient $M_{\alpha}{}^{\beta}$ in the second term and the divergence 
in the last term all produce terms that are proportional to the variable $\Gamma^{\epsilon}_{\kappa\lambda}$. Since none of the rewritten terms can depend explicitly on this variable in the new covariant arguments, we must conclude that
\begin{equation*}
0=2 \frac{\partial C}{\partial  g_{\rho;\gamma\mu(\beta_1|}}\, M_{\rho}{}^{|\beta_2}\delta^{\beta_3)}_{\epsilon} \delta^\kappa_{\gamma}\delta^{\lambda}_{\mu}-2 \frac{\partial C}{\partial  g_{\rho;\gamma(\beta_1\beta_2|}}\,\delta^{|\beta_3)}_\epsilon\,M_{\rho}{}^{\nu} \delta^{(\kappa}_{\nu}\delta^{\lambda)}_{\gamma}\,.
\end{equation*}
Contracting the indices $\epsilon$ and $\kappa$ then leads to the equation
\begin{equation*}
0=\frac{\partial C}{\partial  g_{\rho;\lambda(\beta_1\beta_2|}} \, M_{\rho}{}^{|\beta_3)}-\frac{\partial C}{\partial  g_{\rho;\beta_1\beta_2\beta_3}} \, M_\rho{}^{\lambda}\,.
\end{equation*}
The same logic can now be applied to master equation (5). This time, however, rewriting this equation using the chain rule, and the useful identity above, produces terms which are purely covariant and terms that are proportional to the non-covariant variables $\Gamma^{\alpha}_{(\beta\mu,\nu)}$ as well as terms that are quadratic in $\Gamma^{\alpha}_{\beta\gamma}$. Again the latter must vanish individually. Carefully extracting all information that can be deduced from the vanishing of these terms one finds that
\begin{equation*}
0=\frac{\partial C}{\partial  g_{\rho;\lambda\kappa(\beta_1|}}\,M_\rho{}^{|\beta_2)}\,.
\end{equation*}
The last two equations and the fact that the coefficient $M_{\rho}{}^{\alpha}$ is invertible imply that the potential cannot depend on $ g_{\rho;\alpha\beta\gamma}$. Reducing the master equations (5) and (6) accordingly, one repeats these steps to conclude that the potential $C$ cannot not depend on $ g_{\rho;\alpha\beta}$ either. Now one sees that one may rewrite master equation (4) in the form 
\begin{equation*}
\label{Threeprime}
 0 = (\text{covariant terms}){}^{\beta} + Z^{\mu \nu \beta}{}_{\sigma} \Gamma^{\sigma}_{\mu \nu}
\end{equation*}
for some coefficients $Z^{\mu\nu\beta}{}_{\sigma}$, which however must vanish, since the first part cannot explicitly depend on the variables $\Gamma^{\alpha}_{\beta \gamma}$. Thus also the  partial trace $Z^{\mu\nu\beta}{}_{\mu}$ vanishes, which amounts to
a simple relation for the single potential $C^\mu$ (which is the potential $C_{B}$ with $B = {}^\mu$), 
\begin{equation*}
\label{VanishC}
C^{\alpha} \left(4 \delta^\mu_\alpha  g^\tau  + \delta^{\tau}_\alpha  g^\mu (1- g^\rho  g_\rho) + 7  P^{\mu\tau}  g_\alpha + 9 g_\alpha  g^\mu  g^\tau  \right) =0\,,
\end{equation*}
which upon contraction with $ g_\mu  g_\tau$, reinsertion of the result (namely that $C^\alpha  g_\alpha=0$) into the original equation and a further contraction with $ g_\tau$ yields 
\begin{equation*}
 C^\mu =0\,.
\end{equation*}
But then master equations ($9_N$) and (1) imply that the potential $C$ cannot depend on any of the third derivatives of the fields $P^{\alpha\beta}, g_\alpha, \phi$. Hence, from here on, we can treat the potential $C$ and the tensor potentials on the same footing. 

\paragraph{Vanishing of the potentials $C^\mu{}_{B_1 \dots B_N}$.}  
A similar argument like the one employed before can be applied to master equations ($10_N$) and ($11_N$), whose validity now extends to $N=1$, since the scalar potential $C$ depends on at most second derivatives of the geometric fields. In particular, master equation ($10_N$) can straightforwardly be rewritten in the covariant form
\begin{equation*}
V^{\nu \beta}{}_{B_1\dots B_N}=0\quad\text{for}\quad N\geq 1\,,
\end{equation*}
while master equation $(11_N)$ again contains terms proportional to $\Gamma^{\alpha}_{\beta \gamma}$,
\begin{equation*}
 0 = (\text{covariant terms}){}_{B_1\dots B_N}^{\beta} + Z^{\mu \nu \beta}{}_{\sigma\,B_1\dots B_N} \Gamma^{\sigma}_{\mu \nu} \quad \text{for} \quad N\geq 1 
\end{equation*}
for some coefficients $Z^{\mu \nu \beta}{}_{\sigma\,B_1\dots B_N}$, which must again vanish, and with them the partial trace $Z^{\mu \nu \beta}{}_{\mu\,B_1\dots B_N}$. Suprisingly, taking the difference of the two tensorial quantities $V^{\nu \beta}{}_{B_1\dots B_N}$ and $Z^{\mu \nu \beta}{}_{\mu\,B_1\dots B_N}$ for the same $N$ yields
\begin{equation*}
C^{\alpha}{}_{B_1\dots B_N} \left(4 \delta^\nu_\alpha  g^\beta  + \delta^{\beta}_\alpha  g^\nu (1- g^\rho  g_\rho) + 7  P^{\nu\beta}  g_\alpha + 9 g_\alpha  g^\nu  g^\beta  \right) =0 \qquad \text{for} \quad N\geq 1\,,
\end{equation*}
and, thus, using the same argument as above, we conclude that
\begin{equation*}
C^{\mu}{}_{B_1\dots B_N}=0 \quad \text{for}\quad N\geq1.
\end{equation*}
This is an important result that simplifies the master equations considerably. We have thus learned, in combination with the previous paragraph, that all potentials $C_{B_1 \dots B_N}$ for which at least one of the capital indices takes the value `${}^\alpha$' vanish. In other words, the series expansion of the Lagrangian (see Step 7 of the general recipe) cannot contain any of the velocities $ K_{\alpha}$ belonging to the variable $g_\alpha$.

\paragraph{Potentials do not depend on derivatives of $g_\alpha$.}
Finally we switch back, for a moment, to the master equations as expressed in the partial, rather than covariant, derivatives of the geometric hypersurface fields and show that the remaining potentials $C_{B_1\dots B_N}$ with ${}_{B_i}=({}_{\alpha \beta},\,{}_0)$ and the potential $C$ cannot depend on the first and second partial derivatives of the variable $ g_\alpha$ at all. Setting ${}_{B_N}={}^\rho$ in the symmetry condition ($9_N$), we learn that none of the potentials $C_{B_1\dots B_N}$ (for $N\geq1$)  can depend on $ g_{\alpha,\beta \gamma}$. For the potential $C$, we already concluded this from the master equations ($5$) and ($6$). Thus, the second partial derivatives of $g_\alpha$ cannot appear in any of the potentials. The same holds true for the first partial derivatives $ g_{\alpha,\beta}$. This can be seen from master equations ($11_N$) and ($3$) setting ${}_{B_1}={}^\rho$, which yields
\begin{equation*}
\frac{\partial C_{B_2\dots B_N}}{\partial  g_{\rho,\alpha}}=0 \quad \text{for} \quad N\geq 1\,.
\end{equation*}
Finally, we can even show that potentials $C_{B_1\dots B_N}$ for which at least one of the capital indices is the symmetric pair $\alpha\beta$, cannot depend on the variable $ g_\alpha$ at all. Writing out the divergence in master equations ($11_N$) and ($3$), and using the fact that now nothing in both equations depends on $ g_{\alpha,\gamma}$, we obtain
\begin{equation*}
\frac{\partial^2 C_{B_1\dots B_{N-1}}}{\partial  G^{B_N}{}_{,\alpha \gamma} \partial  g_\mu} =0 \quad \text{for} \quad N\geq 1.
\end{equation*}
This result can be used right away when taking the derivative of master equations ($10_N$) and ($2$) with respect to $g_\sigma$, noticing that we can invert the coefficient $U^{\alpha \beta\,\mu\nu}$. This yields
\begin{equation*}
\label{notong}
\frac{\partial C_{B_1 \dots B_N}}{\partial  g_{\sigma}} = 0 \quad \textrm{if at least one } {}_{B_i}={}_{\alpha \beta}\,
\end{equation*}
at first for any $N\geq 2$, which however can be extended to hold for $N\geq 1$, as one see by evaluating the divergence in the first term in equation ($4$). Thus none of the potentials $C_{B_1\dots B_N}$ (for $N\geq1$ and some  ${}_{B_i}={}_{\alpha \beta}$) depends on $ g_\alpha$. It is, however, not possible to extend this result to all potentials. The potentials $C_{0\dots0}$ (where all capital indices take the value `$0$') and the potential $C$ can still depend on $ g_\alpha$.

\paragraph{Maximally simplified master equations.} 
Taking all of the above findings into account, the remaining master equations (with all others being identically satisfied) are 

\begin{tabular}{ll}
($4''$)\qquad\qquad\qquad &  $0= \nabla_{\mu}(C_{\rho \sigma} \,U^{\rho\sigma\,\beta\mu}) - C_0 \,\nabla^{\beta} \phi + \frac{\partial C}{\partial g_{\rho}} \,M{}_{\rho}{}^\beta\,,$ \\
($9''_{N\geq 2}$) \qquad\qquad\qquad & $\frac{\partial C_{B_1\dots B_{N-1}}}{\partial  G^{B_N}{}_{,\gamma \delta}} = \frac{\partial C_{(B_1\dots B_{N-1}|}}{\partial  G^{|B_N)}{}_{,\gamma \delta}} \,,$\\
($10''_ {N\geq 1}$) \qquad\qquad\qquad & $0 = (N+1)! C_{\mu\nu B_1\dots B_N} \,U^{\mu\nu\,\alpha \beta}- (N-2)(N-1)!\,  \frac{\partial C_{B_1\dots B_{N-1}}}{\partial  G^{B_N}{}_{,\alpha \beta}}\,,$\\
($11''_{N\geq 1}$) \qquad\qquad\qquad & $0 = - (N+1)! \,C_{0 B_1\dots B_N} \nabla^{\beta}\phi + q (N-1)!\, \frac{\partial C_{B_1\dots B_{q-1} B_{q+1}\dots B_N}}{\partial  \phi_{;\beta}}$\\
& $- (N-q)(N-1)!\, \frac{\partial C_{B_1\dots B_q (B_{q+1} \dots B_{N-1}|}}{\partial \phi_{;\rho\sigma}}\, {}^{ P}\Gamma^{\tau \beta}{}_{\rho \sigma |B_N)} \phi_{;\tau}$ \\
& $- 2(N-1)!\, \nabla_{\gamma}\frac{\partial C_{B_1\dots B_{N-1}}}{\partial G^{B_N}{}_{,\gamma\beta}} + N!\,\frac{\partial C_{B_1 \dots B_{N}}}{\partial  g_{\rho}} M{}_{\rho}{}^\beta \,,$
\end{tabular}

where the indicator $q$ denotes the number of capital indices taking the value `$0$', whereas $N-q$ is the number of capital indices $B_i$ being symmetric pairs `$\alpha_i\beta_i$', and the coefficients ${}^{ P}\Gamma$ are defined in (\ref{GammaCoeff}).  In order to not make the equations appear too complicated, we have not written out the chain rule for derivatives with respect to the second partial derivatives of the fields $ G^A$. 

\item[Step 8. Covariance equations] 
Since the scalar potential $C$ can depend on at most the second partial derivatives of the fields, exactly like the the tensor potentials $C_{B_1 \dots B_N}$, the covariance equations take the same form for all $N\geq 0$. The first covariance equation (obtained by differentiation with respect to the second derivatives of the Jacobian) reads
$$0= 2  P^{\mu (\alpha|}\, \frac{\partial C_{B_1\dots B_N}}{\partial  P^{\mu \rho}{}_{,|\beta\gamma)}} -  g_\rho\, \frac{\partial C_{B_1\dots B_N}}{\partial  g_{(\alpha,\beta \gamma)}}\,,$$
while the second one (obtained by differentiation with respect to the first derivatives of the Jacobian) takes the form 
\begin{eqnarray*}
0 = &&2\,  P^{\mu (\alpha|} \, \frac{\partial C_{B_1\dots B_N}}{\partial  P^{\mu \rho}{}_{,|\beta)}}+ 4  P^{\mu (\alpha|}{}_{,\nu} \,\frac{\partial C_{B_1\dots B_N}}{\partial  P^{\mu \rho}{}_{,|\beta)\nu}}-  P^{\mu \nu}{}_{,\rho} \,\frac{\partial C_{B_1\dots B_N}}{\partial  P^{\mu \nu}{}_{,\alpha \beta}}\nonumber\\
&& -  g_\rho\, \frac{\partial C_{B_1\dots B_N}}{\partial  g_{(\alpha,\beta)}} -  g_{\mu,\rho} \,\frac{\partial C_{B_1\dots B_N}}{\partial  g_{\mu,\alpha\beta}} - 2 \, g_{\rho, \mu} \, \frac{\partial C_{B_1\dots B_N}}{\partial  g_{(\alpha,\beta)\mu}} -  \phi_{,\rho} \, \frac{\partial C_{B_1\dots B_N}}{\partial  \phi_{,\alpha \beta}}\,.
\end{eqnarray*}
The third covariance equation (obtained by differentiation with respect to the Jacobian) will not be needed. 

\item[Step 9. Solution of the master and covariance equations]
When solving the master equations arrived at in Step 7, we have to keep in mind that only the potentials $C_{0\dots0}$ and the potential $C$ may depend on the variable $ g_\alpha$. In general, all unknowns $C_{B_1\dots B_N}$ can, in addition, only depend on the variables $( P^{\alpha\beta}, R_{\alpha\beta}, \phi, \phi_{;\alpha}, \phi_{;\alpha\beta})$ because of the covariance equations (\ref{invarS}) and (\ref{invarGamma}), and we already used the fact that the Riemann tensor in three dimensions can be expressed by the Ricci tensor $R_{\alpha\beta}$. 

It is a general result that the potentials $C_{B_1\dots B_N}$ for $N\geq 1$ can depend on the second derivatives of the fields $ G^A$ only up to cubic order \cite{WittePhD}, and since here additionally the second derivative of the scalar field $ \phi$ does not appear in the first covariance equation obtained in Step 8, we can conclude that the Ricci tensor $R_{\alpha\beta}$ can only appear linearly. Moreover, mixed terms, which contain the second derivatives of $ \phi$ and the Ricci tensor, can only be linear in both, as one observes by combining the first covariance equation in the original arguments with the symmetry condition one obtains from writing out the divergence term in equation ($11''_N$) as the
\begin{equation*}
\textrm{symmetry condition}\qquad\frac{\partial^2 C_{B_1\dots B_N}}{\partial  G^{M}{}_{,\alpha (\beta} \partial  G^N{}_{,\gamma\delta)}} \quad\text{for}\quad N\geq0\,.
\end{equation*}

Next, we derive an equation that only involves the potential $C$. To this end, we consider the master equation ($10''_N$) for $N=1$ and $q=1$, and solve it for the potential $C_{\alpha \beta\, 0}$, which yields
\begin{equation*}
C_{\alpha \beta\, 0} = \frac{1}{4}  P_{\gamma (\alpha}  P_{\beta)\delta} \frac{\partial C}{\partial  \phi_{;\gamma \delta}}\,. 
\end{equation*}
On the other hand, considering equation ($11''_N$) for $N=1$ and $q=0$, we have that
\begin{equation*}
0= 2 C_{\alpha\beta \,0} \nabla^{\beta}  \phi - \frac{\partial C}{\partial \phi_{;\rho\sigma}} {}^{ P}\Gamma^{\tau \gamma}{}_{\rho \sigma \alpha \beta} \phi_{;\tau} - 2 \nabla_{\gamma}\frac{\partial C}{\partial  P^{\alpha \beta}{}_{,\mu\beta}}\,,
\end{equation*}
because $C_{\alpha\beta}$ does not depend on $ g_\rho$. Combining both equations, using the explicit form (\ref{GammaCoeff}) of ${}^{ P}\Gamma$, we obtain
\begin{equation*}
0= \frac{\partial C}{\partial  \phi_{;\rho\sigma}} \left(\delta^{\tau}_{(\rho}  P_{\sigma)(\alpha}\delta^{\gamma}_{\beta)} - P_{\rho(\alpha} P_{\beta) \sigma}  P^{\tau \gamma}\right)\nabla_{\tau}\phi - 2 \nabla_{\mu}\frac{\partial C}{\partial P^{\alpha \beta}{}_{,\mu\gamma}}\qquad\qquad(*)\,
\end{equation*}
which constrains the dependence of the potential $C$ on the second derivatives of the fields $ P$ and $ \phi$. Knowing the polynomial dependencies of the potential $C$ on the second derivatives of $ \phi$ and the Ricci tensor $R_{\alpha\beta}$, we may now derive the form of the terms that contain the latter. First, we observe that because of the symmetry condition displayed further above, the last term of the previous equation drops out, as one sees by expanding the divergence 
\begin{equation*}
\nabla_{\mu} \frac{\partial C}{\partial  P^{\alpha\beta}{}_{,\mu \gamma}}= \frac{\partial^2 C}{\partial  \phi\,\partial   P^{\alpha\beta}{}_{,\mu \gamma}}\nabla_\mu\phi + \frac{\partial^2 C}{\partial  \phi_{;\rho}\,\partial   P^{\alpha\beta}{}_{,\mu \gamma}}\nabla_\mu \nabla_{\rho} \phi + \frac{\partial^2 C}{\partial  \phi_{;\rho \sigma}\,\partial   P^{\alpha\beta}{}_{,\mu \gamma}}\nabla_\mu \nabla_{\rho}\nabla_\sigma \phi\,
\end{equation*}
and rewriting
\begin{eqnarray*}
\nabla_\mu\nabla_\rho\nabla_\sigma \phi&=&-\frac{2}{3} \left(  P_{\nu[\mu}\delta^{(\kappa}_{\rho]} \delta^{\tau)}_{\sigma}-  P_{\sigma[\mu}\delta^{(\kappa}_{\rho]} \delta^{\tau)}_{\nu} +  P_{\nu[\mu}\delta^{(\kappa}_{\sigma]} \delta^{\tau)}_{\rho}-  P_{\rho[\mu}\delta^{(\kappa}_{\sigma]} \delta^{\tau)}_{\nu} \right) R_{\kappa\tau} \nabla^{\nu} \phi \nonumber\\
&&+\frac{1}{3} ( P_{\nu[\mu} P_{\rho]\sigma} +  P_{\nu[\mu} P_{\sigma]\rho})  P^{\kappa \tau} R_{\kappa\tau} \nabla^{\nu}  \phi + \nabla_{(\mu}\nabla_\rho\nabla_{\sigma)} \phi\,.
\end{eqnarray*}
We can then use the resulting equation to compare the different powers of the second derivatives of $ \phi$ and the Ricci tensor $R_{\alpha\beta}$ appearing in the potential $C$. Note that none of these terms can depend explicitly on $ g_\alpha$, because of the second last equation derived in paragraph {\it c.} of Step 7 above, which simplifies matters significantly. It follows, for example, that the coefficient in the cubic part $C_{\text{cubic}}^{\rho\sigma\mu\nu\kappa\epsilon}\,  \phi_{;\rho\sigma}\,  \phi_{;\mu\nu}\, \phi_{;\kappa\epsilon}$ of $C$ has to satisfy
\begin{equation*}
0=  C_{\text{cubic}}^{\rho\sigma\mu\nu\kappa\epsilon} \left(\delta^{\tau}_{(\rho}  P_{\sigma)(\alpha}\delta^{\gamma}_{\beta)} -  P_{\rho(\alpha} P_{\beta) \sigma}  P^{\tau \gamma}\right)\nabla_{\tau} \phi\,.
\end{equation*}
However, it is easy to see that the term in brackets can be inverted, which implies that there cannot be such a cubic term in $C$. For the mixed term $C_{\text{mixed}}^{\alpha\beta\gamma\delta}\, R_{\alpha\beta}\;  \phi_{;\gamma\delta}$, only the last term in $(*)$ is relevant. A brute-force calculation then shows that also this term has to vanish. 

The remaining terms can then be investigated by making the exhaustive ansatz
\begin{eqnarray*}
C&=&\sqrt{-\det P_{\alpha\beta}}\Big[C_f( \phi,\nabla_\alpha  \phi \nabla^\alpha  \phi, g^\alpha\nabla_\alpha  \phi , g^\alpha g_{\alpha})+ R_{\alpha\beta} (a_1\,  P^{\alpha\beta}+ a_2\, \nabla^\alpha \phi\nabla^\beta \phi)\nonumber\\
&& \quad\qquad\qquad+ \nabla_\alpha\nabla_\beta \phi (a_3\,  P^{\alpha\beta}+ a_4\, \nabla^\alpha \phi\nabla^\beta \phi) +\nabla_\alpha\nabla_\beta \phi\nabla_\gamma\nabla_\delta \phi \,(a_5\,  P^{\alpha\beta} P^{\gamma\delta}\nonumber\\
&&\quad\qquad\qquad+a_6 \, P^{\alpha\gamma}  P^{\beta\delta} + a_7\,  P^{\alpha\beta}\nabla^\gamma \phi\nabla^\delta \phi+a_8\, P^{\alpha\gamma}\nabla^\beta \phi\nabla^\delta \phi\nonumber\\
&&\quad\qquad\qquad+a_9\,\nabla^\alpha \phi\nabla^\beta \phi\nabla^\gamma \phi\nabla^\delta \phi)\Big]\,
\end{eqnarray*}
where the scalar functions $a_i$ may depend on $ \phi$ and $\nabla_\alpha\phi\nabla^{\alpha} \phi$ and the free function $C_f$ depends on all scalars indicated in brackets. Thus extracting all information in equation $(*)$, one is led to a system of linear differential equations for the functions $a_i$, which can be solved uniquely to yield the most general form of the potential $C$ allowed by the master equations:
\begin{equation*}
C=\sqrt{-\det  P_{\alpha\beta}} \left[ a_1( \phi) R- 2 \frac{da_1( \phi)}{d  \phi}  P^{\alpha\beta}  \phi_{;\alpha\beta} +C_f( \phi,\nabla_\alpha  \phi \nabla^\alpha  \phi, g^\alpha\nabla_\alpha  \phi , g^\alpha g_{\alpha}) \right] \,.
\end{equation*}

A similar procedure can be applied to determine the potential $C_{\rho\sigma}$, which, as we know, cannot depend on $ g_\alpha$. We can even derive two independent equations for $C_{\rho\sigma}$. The first of these is given by equation ($11''_N$) for $N=2$ and $q=0$, i.e., 
\begin{equation*}
0=-\frac{\partial C_{\{\rho\sigma|}}{\partial  \phi_{;\mu\nu}} {}^{ P}\Gamma^{\tau\beta}{}_{\mu\nu|\epsilon\kappa\}} \nabla_\tau  \phi - \nabla_{\gamma}\frac{\partial C_{\rho\sigma}}{\partial  P^{\epsilon\kappa}{}_{,\gamma\beta}}\,,
\end{equation*}
where the symmetrization brackets $\{\dots\}$ are to be understood as symmetrizing the pairs $\rho \sigma$ and $\epsilon\kappa$, but not the individual indices. Here, we made use of the facts that $C_{\alpha\beta \,0}$ does not depend on $ g_\alpha$ either, and that, from equation ($10''_N$) with $N=2$, we may conclude that $C_{\alpha\beta\,B_1 B_2}=0$. The second equation can be derived from equation ($11''_N$) with $N=2$ and $q=1$ using the same reasoning, which leads to
\begin{equation*}
0=\frac{C_{\rho\sigma}}{\partial  \phi_{;\beta}} - \frac{\partial C_{0}}{\partial  \phi_{;\mu\nu}} {}^{ P}\Gamma^{\tau\beta}{}_{\mu\nu\rho\sigma} \nabla_\tau  \phi  - \nabla_\gamma \frac{\partial C_{\rho\sigma}}{\partial  \phi_{;\gamma\beta}}\,,
\end{equation*}
where we have already used the master equation ($9''_N$) in the last term. The potential $C_0$, which still appears in this equation, can be eliminated by solving equation ($4''$), so that 
\begin{equation*}
C_0 = \frac{1}{\nabla_\rho  \phi \nabla^{\rho} \phi} \left[ \nabla_\beta \phi \nabla_{\mu} (C_{\kappa\tau} U^{\kappa \tau\, \beta\mu}) + \frac{\partial \tilde C_f}{\partial  g_{\rho}} M_{\rho}{}^{\beta} \nabla_\beta  \phi\right]\,
\end{equation*}
with $\tilde C_f=\sqrt{-\det P_{\alpha\beta}}\, C_f$. Inserting this back into the second last equation, the second term in brackets vanishes because of the most general form for the scalar potential obtained above, and hence we obtain
\begin{equation*}
0= \frac{C_{\rho\sigma}}{\partial  \phi_{;\beta}} - \frac{1}{\nabla_\rho  \phi \nabla^{\rho}  \phi}\,  {}^{ P}\Gamma^{\tau\beta}{}_{\mu\nu\rho\sigma}\,U^{\xi\delta\,\psi\zeta}\, \nabla_\tau  \phi  \nabla_\psi  \phi \nabla_{\zeta}C_{\xi\delta}- \nabla_\gamma \frac{\partial C_{\rho\sigma}}{\partial  \phi_{;\gamma\beta}}\,.
\end{equation*}
Using equation the above two equations for the potential $C_{\rho\sigma}$, we can now constrain the form of the latter the same way we did for the potential $C$. First of all, writing out the divergence in equation ($4''$), one can conclude that $C_{\rho\sigma}$ can be at most linear in $R_{\alpha\beta}$ and at most quadratic in $\phi_{;\alpha\beta}$. This is the case because the resulting symmetry condition also involves the symmetric pair of indices of $C_{\rho\sigma}$, and, thus, strengthens the two symmetry conditions we already used for the potential $C$. There cannot be any terms mixing $R_{\alpha\beta}$ and  $\phi_{;\alpha\beta}$ for the same reason. Evaluating all information contained in the two equations  for $C_{\rho\sigma}$, one obtains, as a preliminary result, that
\begin{eqnarray*}
C_{\rho\sigma}&=&\sqrt{-\det  P_{\alpha\beta}} \Big[(b_1  \phi + b_2) (R_{\rho\sigma} - \frac{1}{2}  P_{\rho\sigma} R) + b_3\, R\, P_{\rho\sigma} \nonumber\\
&&\qquad\qquad\qquad+ \frac{1}{2} b_1 ( P^{\alpha\beta} \phi_{;\alpha\beta}\, P_{\rho\sigma} - \phi_{;\rho\sigma}) + a_2( \phi)  P_{\rho\sigma}\Big]\,,
\end{eqnarray*}
with constants $b_1,b_2,b_3$ and a new unknown function $a_2( \phi)$. From the above expression for the potential $C_0$, however, we can then directly conclude that $b_3=0$, since this equation cannot contain third partial derivatives of $ P^{\alpha\beta}$; a straightforward calculation yields
\begin{eqnarray*}
C_0&=&\sqrt{-\det  P_{\alpha\beta}} \Big[-\frac{b_1}{\nabla_\rho  \phi\nabla^\rho  \phi} \nabla^{\alpha}   \phi \nabla^\beta  \phi (R_{\alpha\beta} -   P_{\alpha\beta} R) \nonumber\\
&&\qquad\qquad\qquad- 2 \frac{da_2(  \phi)}{d  \phi}+ \frac{1}{\nabla^\alpha  \phi\nabla_\alpha  \phi}\frac{\partial C_f}{\partial   g_{\rho}} M_{\rho}{}^{\beta} \nabla_\beta  \phi\Big]\,.
\end{eqnarray*}
Now consider equation ($11''_N$) for $N=2$ and $q=2$, which amounts to
\begin{equation*}
0= -3! \, C_{000} \nabla^\beta\phi+2! \frac{\partial C_{00}}{\partial  g_\rho}\, M_{\rho}{}^{\beta} + 2 \frac{\partial C_{0}}{\partial \phi_{;\beta}}\,.
\end{equation*}
Since we know that $C_{\alpha\beta\,00}=0$, the master equation ($9''_N$) implies that the potential $C_{000}$ cannot depend on $R_{\rho\sigma}$. Moreover, since $\partial C_{00}/\partial g_\rho$ cannot contain $R_{\rho\sigma}$ either, the last equation implies that $b_1=0$. Thus, we arrive at
\begin{eqnarray*}
C_{\rho\sigma}&=&\sqrt{-\det P_{\alpha\beta}}\left[b_2 (R_{\rho\sigma} - \frac{1}{2}  P_{\rho\sigma} R)+ a_2( \phi) P_{\rho\sigma} \right]\quad\text{and}\\
C_0&=&\sqrt{-\det P_{\alpha\beta}}\,\left[-2 \frac{da_2(\phi)}{d\phi}+ \frac{1}{\nabla_\alpha\phi \nabla^\alpha \phi} \frac{\partial  C_f}{\partial  g_{\rho}} M_{\rho}{}^{\beta} \nabla_\beta \phi \right]\,.
\end{eqnarray*}
We can now determine the remaining potentials recursively. Using equation the second equation derived in Step 9, we get
\begin{equation*}
C_{\alpha\beta\, 0}= -\sqrt{-\det P_{\alpha\beta}}\,\frac{1}{2}\frac{da_1(\phi)}{d\phi} P_{\alpha\beta}\,.
\end{equation*}
From equation ($10''_N$) with $N=1$ and $q=0$, we then find the potential 
\begin{equation*}
 C_{\alpha \beta \gamma \delta} = \frac{1}{8}\left[  P_{\alpha \gamma} P_{\beta \delta} +  P_{\beta \gamma} P_{\alpha \delta} - 2  P_{\alpha\beta} P_{\gamma\delta}\right]\,.
\end{equation*}
It is then clear that all other potentials containing at least one index pair $\alpha\beta$ vanish. This can be seen recursively from equation ($10''_N$) and the fact that all potentials with more than two capital indices do not depend on second derivatives of the fields.

Thus, only the potentials with `$0$' indices remain to be determined. Denoting the potentials
$$C_{(N)} := C_{\!\!\underbrace{\textrm{\small 0\dots0}}_{N \textrm{ zeroes }}} \qquad \textrm{ for } N\geq 1$$
and using equation ($11''_N$), we get the
\begin{equation*}
\label{recursion}
\textrm{recursion} \qquad C_{(N+1)}=  \frac{1}{\nabla_\rho \phi \nabla^\rho  \phi}\frac{N!}{(N+1)!}\left[\frac{\partial C_{(N)}}{\partial  g_\gamma} \, M_{\gamma}{}^{\beta} \nabla_{\beta}   \phi+  \frac{\partial C_{(N-1)}}{\partial \nabla_\beta \phi} \nabla_\beta  \phi\right]\,.
\end{equation*} 
for all potentials $C_{(N+1)}$ with $N \geq 1$.

One thus obtains (omitting two additional summands linear in the velocities $K^{\alpha\beta}$ and $K$, which have no impact on the resulting equations of motion) the most general gravitational Lagrangian that can underlie the $SO(p,q)$-violating Dirac dynamics ,
\begin{eqnarray*}
 L&=& \sqrt{-\det  P_{\alpha \beta}}  \Big[2 \frac{d^2 a_1( \phi)}{d \phi^2}  K^2 - \frac{1}{2} \frac{d a_1( \phi)}{d  \phi}\,  K\,  P_{\alpha \beta}  K^{\alpha \beta} \nonumber \\
&&\qquad\qquad\quad- a_1( \phi) \,C_{\alpha \beta \gamma \delta}  K^{\alpha \beta}  K^{\gamma \delta} + a_1( \phi) R - 2 \frac{d a_1( \phi)}{d \phi}  P^{\alpha \beta} \nabla_\alpha \nabla_\beta \phi\nonumber\\
&& \qquad\qquad\quad+\sum_{N=1}^\infty C_{(N)}  K^N + C_{(0)}( \phi,\nabla^{\alpha} \phi\nabla_\alpha \phi,  g^\alpha\nabla_\alpha  \phi, g_\alpha  g^\alpha)\Big]\,,
\end{eqnarray*}
with a freely specifiable function $a_1( \phi)$ (mediating the derivative coupling between the scalar field $\phi$ and the metric $ P^{\alpha \beta}$---a non-derivative coupling thus obviously requires $a_1( \phi)=\text{const}$) and a freely specifiable function $C_{(0)}(\phi, \nabla^\alpha  \phi\nabla_\alpha  \phi, g^\alpha \nabla_\alpha \phi, g_\alpha  g^\alpha)$, in terms of which, however, all potentials $C_{(N)}$ are determined by virtue of the 
\begin{equation*}
\textrm{recursion start} \qquad C_{(1)}=  \frac{1}{\nabla^\alpha \phi\nabla_\alpha \phi}\,\frac{\partial C_{(0)}}{\partial  g_{\rho}} M_{\rho}{}^{\beta} \nabla_\beta  \phi
\end{equation*}
and the recursion formula further above.

A striking feature of the above dynamics is that while the field $g_\alpha$ appears in the potentials $C_{(N)}$, for $N\geq0$, the corresponding velocity $ K_{\alpha}$ does not appear in the Lagrangian at all. But although the geometric field $ g_\alpha$ thus does not have its own `dynamical' equations of motion, it can nevertheless be fully determined by the dynamics of the other variables $ P^{\alpha\beta}$ and $ \phi$ once the recursion is employed. Indeed, in appendix \ref{sec:concretemodel}, we will illustrate this mechanism explicitly, in order to show that the absence of velocity terms $K^\alpha$ does {\it not} imply dynamically undeteremined geometrical degrees of freedom. 

\item[Step 10. Additional energy or symmetry conditions] were not needed to obtain an analytic solution of the master equations in this case. 

\end{description}

In summary, we found the complete family of canonical gravitational dynamics for the vector-tensorial spacetime geometry defined by the metric $g$ and the vector field $W$ that can support the deformed Dirac equation we started from, and which indeed presented our example for a decidely non-standard model type matter action in the introduction. We emphasize again that we did not propose these specific matter equations as phenomenologically relevant matter dynamics, but as an instructive example that shows how to proceed for any matter dynamics the reader may wish to consider for her own phenomenological or theoretical reasons. 

Kinematically, we found that, in this specific case, predictivity and quantizability of the matter field equations amount to the condition that depending on the vector field, the metric part of the tensor-vector geometry may have either Lorentzian or Riemannian signature, with the resulting $SO(1,3)$ or $SO(4)$ symmetry however being directly broken by the vector field part of the geometry. More precisely, if the vector field has $g$-norm less than $-1$, the metric must have Riemannian signature in order to render the matter theory predictive and quantizable, whereas a Lorentzian signature of the metric is enforced in all other cases. While a Riemannian signature for the metric may appear non-physical, it should be noted that this is not the case, since it is the hyperbolicity of the principal tensor that is physically relevant, and that the intuition that the metric should have Lorentzian signature merely stems from the case of Maxwell theory, where the principal tensor indeed is identical with the (inverse) metric, and where this intuition is therefore correct. But only there. 

The comparatively high effort required to solve the master equations for the matter dynamics considered in this second case study indicates how hard it is, in general, to construct an appropriate kinematical and dynamical theory of spacetime that can underpin specific phenomenological models of matter. But at the same time, we saw that it can be done. The complexity of the gravitational Lagrangian obtained in this case further makes it pretty obvious how hopeless it would be to try to arrive at appropriate gravitational dynamics by mere {\it guessing}, without having constructed the pertinent master equations.

\newpage
\section{Conclusions}

Any set of matter field equations---whether considered for phenomenological reasons,  theoretical considerations, or the mere heck of it---must be supplemented by dynamics for their coefficients in order to be completed into a closed theory. Physically, we like to call the degrees of freedom making up the coefficients of matter field equations the geometry of spacetime, and then refer to the dynamics of these degrees of freedom as gravitational dynamics. Using this parlance, in this paper we presented the ten-step recipe for the practical derivation of  gravitational dynamics---namely the derivation of the gravitational Lagrangian as the solution of a set of master equations, which in turn are constructed directly from prescribed matter field dynamics---which underpin the matter field equations of choice such that the latter can be both predictive and quantizable. From this point of view, gravity emerges as a mere auxiliary science. 

The general recipe for the extraction of these master equations from the matter field dynamics comes as ten straightforward rules, and presents the remarkably simple practical essence of a number of combined results, whose conceptual spirit is that of geo\-metrodynamics developed more than five decades ago but whose technical derivation in the broad context considered here required several pieces of decidely more modern mathematical machinery. Now the central point of the present paper is that, once the rules are derived, their {\it application} to concrete matter models no longer requires any more sophisticated mathematical techniques than those taught in any introductory course on general relativity. 

We then demonstrated the concrete application of this so properly founded recipe to two completely worked, instructive case studies. The first one considered Maxwell matter, but  goes through in completely unaltered fashion for any standard model matter dynamics and yields, as the unique solution to the master equations, the Einstein-Hilbert action with a cosmological term. The second case study then considered a particular example of a matter model beyond the standard model, for which we also constructed and then solved the master equations explicitly and thus derived the appropriate gravitational dynamics. By these examples we were able to show, in technical detail, what is needed on the gravitational side in order to make a given linear matter model work. All one has to do is to determine suitable underlying gravitational dynamics according to the general rules we provided.  Given that only about four percent of the matter-energy in the universe appears to be of standard model origin, having such a recipe at one's disposal is hardly a luxury.  

\newpage
The scope of the recipe given here is not restricted to field theoric matter. For one may, instead, start from a particular dispersion relation for massive or massless point matter. Remarkably, it turns out that in order for such dispersion relations to arise as a primary constraint from some point particle action, they must have a covariant formulation in terms of an again bi-hyperbolic tensor which must then be used in lieu of the principal tensor one derives for field matter, and which consequently doubles as both the principal tensor and the fundamental geometric tensor, at which point the recipe can be applied to extract the associated master equations, see \cite{GSWW}. Thus by a different physical mechanism than in the case of field matter, but with precisely the same physical inevitability and the same central technical condition of bi-hyperbolicity, any postulated dispersion relation for point particle matter is suitably constrained and supplemented with a dynamical law by solving the pertinent master equations. 

A pleasant feature of the presented method to obtain gravitational dynamics from prescribed matter dynamics is that the latter contain the entire physical input into the master equations. In other words, the gravitational theory is precisely as physically relevant as the matter model it is extracted from. In case there are various matter fields whose dynamics do not yield the same principal tensor, the principal tensor of the entire theory is quickly seen to be the product of the principal tensors of the individual theories. Thus the remarkable consequence, and wider lesson, is that any new discovery about matter immediately translates into an appropriate gravity theory. Depending on the newly discovered matter dynamics, this could still be standard general relativity or not. The observed matter, and only the observed matter, suffices as an input and will be the judge.

\acknowledgments
The authors would like to thank Gary Gibbons, Jean-Philippe Uzan, Marcus Werner and Shinji Mukohyama for valuable discussions and in particular Gary Gibbons for the suggestion to extract further information from energy conditions on the matter and Jean-Philippe Uzan for suggesting to condense the various technical results obtained in previous work into a practical recipe. CW acknowledges generous support received from the International Max Planck Research School and the Studienstiftung des deutschen Volkes. FPS would like to thank the Kavli Institute for the Physics and the Mathematics of the Universe for their generous hospitality in Tokyo, where part of this work has been done during two extended stays.  

\newpage
\appendix
\section{\label{sec_covexample} Illustration of the derivation of differential covariance equations}

It suffices to describe the method for one case, which is even simpler than the simplest case that can arise in our context. Assume that there is only one hypersurface field $G_\alpha$ and we are aiming at phrasing the condition for some $(0,2)$-tensor field $C_{\mu\nu}$ to be constructed from only $\partial G$ in terms of a partial differential equation. This of course amounts to the condition that
$$C_{\bar\mu\bar\nu}\left(\frac{\partial y^\alpha}{\partial \bar y^{\bar \alpha}} \frac{\partial}{\partial y^\alpha}  \left(\frac{\partial y^\beta}{\partial \bar y^{\bar \beta}} G_\beta\right)\right) = \frac{\partial y^\mu}{\partial \bar y^{\bar \mu}} \frac{\partial y^\nu}{\partial \bar y^{\bar \nu}} C_{\mu\nu}(\partial_\alpha G_\beta)\,,$$
which simply expresses that the tensor components constructed from the transformed field components are the tensorially transformed components construced from the untransformed field components. 

The first step to convert this algebraic condition on $C_{\mu\nu}$ into two partial differential equations for $C_{\mu\nu}$, is to rewrite the algebraic condition in terms of the Jacobian of the coordinate transformation and all its derivatives, i.e., in our example, in terms of 
$$T^\alpha{}_{\bar\alpha} = \frac{\partial y^\alpha}{\partial \bar y^{\bar \alpha}} \quad \textrm{ and } \quad  T^{\beta}{}_{\bar\alpha\bar\beta} = \frac{\partial^2 y^\beta}{\partial\bar y^{\bar\alpha} \partial\bar y^{\bar \beta}}\,,$$
such that it takes the form
$$C_{\bar\mu\bar\nu}\left(T^\alpha{}_{\bar \alpha} T^\beta{}_{\bar \beta} \frac{\partial}{\partial y^\alpha} G_\beta + T^\beta{}_{\bar\alpha\bar\beta} G_\beta\right) = T^\mu{}_{\bar\mu} T^\nu{}_{\bar\nu} C_{\mu\nu}(\partial_{\alpha} G_\beta)\,.$$
Note that $T^\alpha{}_{\bar\alpha\bar\beta}$ is symmetric in its lower indices due to the Schwarz rule, but only because they refer to the same (the barred) set of coordinates --- if the tensor $C_{\mu\nu}$ depended, other than in our current example, on the first partial derivative of a hypersurface vector field $G^\alpha$, rather than a covector field $G_\alpha$, one could however still arrange for the then appearing derivative of the Jacobian to be with respect to coordinates from the same (then the unbarred) set of coordinates, by inserting appropriate factors of the Jacobian or its inverse; similarly one proceeds where higher than second derivatives appear.      

The second step towards converting the algebraic covariance condition into partial differential equations is to derive the former first with respect to the highest derivative of the Jacobian and to evaluate the result at the identity transformation, and then to repeat this with respect to all lower order derivatives of the Jacobian, up to and including the zeroth derivative, i.e., with respect to the Jacobian itself. For the present case, the derivative with respect to $T^{\sigma}{}_{\bar\rho\bar\sigma}$ yields
$$\partial^{\bar\alpha\bar\beta} C_{\mu\nu} \, \delta^\beta_\sigma \delta^{\bar\rho}_{(\bar\alpha} \delta^{\bar \sigma}_{\bar \beta)} G_\beta = 0$$ 
and the derivative with respect to $T^\sigma{}_{\bar\sigma}$ yields
$$\partial^{\bar\alpha\bar\beta} C_{\bar\mu\bar\nu}(\delta^\alpha_\sigma \delta^{\bar \sigma}_{\bar\alpha} \delta^\beta_{\bar\beta} + \delta^\alpha_{\bar \alpha} \delta^{\beta }_{\sigma} \delta^{\bar \sigma}_{\bar\beta}) \frac{\partial}{\partial y^\alpha} G_\beta = (\delta^{\mu}_{\sigma} \delta^{\bar\sigma}_{\bar\mu} \delta^\nu_{\bar\nu} + \delta^{\mu}_{\bar\mu} \delta^{\nu}_{\sigma} \delta^{\bar\sigma}_{\bar\nu}) C_{\mu\nu}$$
which simplify to
$$\partial^{(\bar\alpha\bar\beta)} C_{\bar\mu\bar\nu} = 0 \qquad \textrm{ and } \qquad \partial^{\bar\sigma\beta} C_{\bar\mu\bar\nu} G_{\beta,\sigma} + \partial^{\alpha\bar\sigma} C_{\bar\mu\bar\nu} G_{\sigma,\alpha} = \delta^{\bar\sigma}_{\bar\mu}C_{\sigma\bar\nu} + \delta^{\bar\sigma}_{\bar\nu}C_{\bar\mu\sigma}\,.$$
These two differential equations encode the entire information about $C_{\mu\nu}$ being a second rank covariant tensor constructed from the first derivatives of a covector field $G_\alpha$. (In this case one can solve the covariance equations all by themselves by first observing that the first covariance condition implies that $C_{\mu\nu}$ at most depends on the antisymmetric part $\partial_{[\alpha} G_{\beta]}$ of $\partial_{\alpha} G_{\beta}$ and then considering the contraction of the second equation with respect to $\bar\sigma$ and $\sigma$, i.e., $\partial^{\bar\sigma\bar\tau} C_{\bar\mu\bar\nu} G_{\bar\tau,\bar\sigma} = C_{\bar\mu\bar\nu}$ which, using the insight from the first covariance equation, becomes $\partial^{[\bar\sigma\bar\tau]} C_{\bar\mu\bar\nu} G_{[\bar\tau,\bar\sigma]} = C_{\bar\mu\bar\nu}$ which yields the final result that $C_{\mu\nu}$ must be proportional to  $\partial_{[\mu} G_{\nu]}$. This is the well-known result that without further structure, the only second rank tensor that can be built from the first derivatives of a covector field is the exterior derivative of the latter.)

\section{Field redefinitions suggested by covariance equations}
\label{sec:Invar}
We now discuss what can be extracted from the covariance equations for the case where one of the geometric hypersurface tensor fields $ G^A$ can be formally employed as a hypersurface metric. As discussed in appendix \ref{sec_covexample}, covariance equations reflect the tensor-density nature of the potentials $C_{B_1 \dots B_N}$ for $N\geq1$, which are functions of the form $C_{B_1\dots B_N}( G^A,\partial  G^A, \partial^2  G^A)$. The partial derivatives of the tensor fields $ G^A$ are of course not tensor fields, and hence the covariance equations encode how those non-tensorial fields have to be combined in order to produce the weight-one tensor densities $C_{B_1 \dots B_N}$.

A fruitful idea is to simplify the covariance equations by replacing the arguments $G^A$, $\partial_\gamma  G^A$, $\partial^2_{\gamma\delta}  G^A$, on which the scalar and tensor potentials depend, by a set of arguments that simplifies the covariance equations. In particular, this is possible if one of the fields $G^A$ can be employed as a hypersurface metric. Thus, let us assume that the hypersurface geometry is only given by an inverse metric, so that $G^A=(P^{\alpha\beta})$. For simplicity, we discuss this particular case first, and then generalize it to all cases where, apart from a hypersurface metric, we have an arbitrary number of additional hypersurface tensor fields, $G^A=(P^{\alpha\beta}, \dots)$. The covariance equations for the simple case are
\begin{equation}\label{invaronedegP2}
0= P^{\alpha (\sigma}\,\frac{\partial C_{B_1\dots B_N}}{\partial\partial^2_{\mu\nu)}  P^{\alpha\rho}}\,
\end{equation}
and
\begin{equation}\label{invartwodegP2}
0=2\, P^{\alpha (\mu}\frac{\partial C_{B_1\dots B_N}}{\partial \partial_{\nu)}  P^{\alpha\rho}}-\partial_\rho  P^{\alpha\beta}\frac{\partial C_{B_1\dots B_N}}{\partial\partial^2_{\mu\nu} P^{\alpha\beta}}+\,4\,\partial_{\sigma} P^{\alpha(\mu}\frac{\partial C_{B_1\dots B_N}}{\partial\partial^2_{\nu)\sigma} P^{\alpha\rho}}\,.
\end{equation}
Since the field $ P^{\alpha \beta}$ can be employed as a hypersurface metric, we can now perform a change of arguments from $( P^{\alpha\beta},\partial_\gamma P^{\alpha\beta},\partial^2_{\gamma\delta} P^{\alpha \beta})$ to a new set of arguments $( P^{\alpha\beta}, \Gamma^{\alpha}_{\beta\gamma}, R_{\alpha\beta\gamma\delta}, S_{\alpha\beta\gamma\delta})$, trading the first partial derivatives of the field $ P^{\alpha\beta}$ for the Levi-Civita connection coefficients $\Gamma$ of $ P^{\alpha\beta}$, and its second partial derivatives for the corresponding Riemann-Christoffel tensor $R$ and another variable $S$. Explicitly this transformation is given by
\begin{eqnarray}
\Gamma^{\alpha}_{\beta \gamma} &=&{}^{ P}\Gamma^{\alpha \rho}{}_{\beta\gamma \lambda \kappa}  P^{\lambda \kappa}{}_{,\rho} \label{Var1}\\
R_{\alpha \beta \gamma \delta}  &=& R_1{}^{\mu \nu}{}_{\kappa \tau \alpha \beta \gamma \delta}\,  P^{\kappa \tau}{}_{,\mu\nu} + R_2{}^{\sigma \tau}{}_{\mu \nu \kappa \epsilon \alpha \beta \gamma \delta}\,  P^{\mu \nu}{}_{,\sigma}  P^{\kappa \epsilon}{}_{,\tau} \label{Var2}\\
S_{\alpha\beta\gamma\delta} &=& S_1{}^{\mu \nu}{}_{\kappa \tau \alpha \beta \gamma \delta}\,  P^{\kappa \tau}{}_{,\mu\nu} + S_2{}^{\sigma \tau}{}_{\mu \nu \kappa \epsilon \alpha \beta \gamma \delta}\,  P^{\mu \nu}{}_{,\sigma}  P^{\kappa \epsilon}{}_{,\tau}\,, \label{Var3}
\end{eqnarray}
where for brevity we used a comma to denote partial derivatives. The coefficients in the above expressions are
\begin{eqnarray}
{}^{ P}\Gamma^{\alpha \rho}{}_{\beta \gamma \kappa \tau} &:=& \frac{1}{2}  P_{\beta(\kappa} P_{\tau)\gamma}  P^{\alpha \rho} - \delta^{\alpha}_{(\kappa}  P_{\tau)(\beta} \delta^{\rho}_{\gamma)}\,,\label{GammaCoeff}\\
R_1{}^{\mu\nu}{}_{\kappa\tau \alpha \beta\gamma \delta}&:=& 2 \delta^{(\mu}_{[\beta}  P_{\alpha](\kappa} P_{\tau)[\gamma}\delta^{\nu)}_{\delta]}\,,\\
R_2{}^{\sigma \tau}{}_{\mu\nu\kappa \epsilon\alpha \beta \gamma \delta}&:=& \delta^{\tau}_{(\nu}  P_{\mu)[\alpha} P_{\beta](\kappa}  P_{\epsilon)[\delta}\delta^\sigma_{\gamma]} + \delta^{\tau}_{(\nu}  P_{\mu)[\delta} P_{\gamma](\kappa}  P_{\epsilon)[\alpha}\delta^\sigma_{\beta]} + \delta^{\sigma}_{[\alpha}  P_{\beta](\kappa} P_{\epsilon)(\mu}  P_{\nu)[\gamma}\delta^\tau_{\delta]}\nonumber\\
&& + 2\,  \delta^{\sigma}_{[\alpha}  P_{\beta](\mu} P_{\nu)(\kappa}  P_{\epsilon)[\gamma}\delta^\tau_{\delta]} +\frac{1}{2}\,  P_{(\mu|[\alpha}  P_{\beta](\kappa} P_{\epsilon)|\nu)} \delta^{\sigma}_{[\delta}\delta^{\tau}_{\gamma]}\nonumber\\
&& + \frac{1}{2}\,  P_{(\mu|[\delta}  P_{\gamma](\kappa} P_{\epsilon)|\nu)} \delta^{\sigma}_{[\alpha}\delta^{\tau}_{\beta]} + \frac{1}{2}\,  P^{\sigma \tau}  P_{(\mu|[\alpha}  P_{\beta](\kappa} P_{\epsilon)[\gamma}  P_{\delta]|\nu)}\,,\\
S_1{}^{\mu\nu}{}_{\kappa\tau\alpha\beta\gamma\delta}&:=& -  P_{\alpha(\kappa} P_{\tau)(\beta} \delta^{(\mu}_{\gamma} \delta^{\nu)}_{\delta)}+\frac{1}{2}  P_{(\kappa|(\beta}\delta^{(\mu}_{\gamma} P_{\delta)|\tau)} \delta^{\nu)}_{\alpha}\quad\text{and}\\
S_2{}^{\sigma \tau}{}_{\mu\nu\kappa\epsilon\alpha\beta\gamma\delta}&:=& 2  P_{\alpha(\mu} P_{\kappa)(\beta} \delta^\sigma_{\gamma} \delta^{\tau}_{\delta)}  P_{\nu\epsilon} -  P_{(\beta|(\mu} P_{\kappa)|\gamma} \delta^{\sigma}_{\delta)}  P_{\nu\epsilon} \delta^{\tau}_\alpha\,.
\end{eqnarray}
The variable $S_{\alpha\beta\gamma\delta}$ is needed since the Riemann tensor does not contain all the second partial derivatives of the field $ P^{\alpha\beta}$. Without this variable, the change of arguments is not invertible. We note that the variables $S_{\alpha\beta\gamma\delta}$ are not components of a tensor and feature the symmetry $S_{\alpha\beta\gamma\delta}=S_{\alpha(\beta\gamma\delta)}$. In order to express the original covariance equations now with respect to the new arguments, we also need the inverse transformation:
\begin{eqnarray}
  P^{\alpha \beta}{}_{,\gamma}&=& -2  P^{\mu (\alpha} \Gamma^{\beta)}_{\mu\gamma}\label{inverseVar1}\\
 P^{\mu\nu}{}_{,\gamma\delta}&=&\frac{1}{3}  P^{\mu \alpha}  P^{\nu\beta} \left(R_{\alpha\gamma\beta\delta} + R_{\beta\gamma\alpha\delta}\right) -  P^{\mu \alpha}  P^{\nu\beta}\left(S_{\alpha\beta\gamma\delta}+ S_{\beta\alpha\gamma\delta}\right) \nonumber\\
&& +\frac{1}{3}  P_{\rho\sigma}  P^{\mu\alpha} P^{\nu\beta} \left(\Gamma^{\rho}_{\beta(\gamma}\Gamma^{\sigma}_{\delta)\alpha} + 2 \Gamma^{\rho}_{\gamma\delta} \Gamma^{\sigma}_{\alpha\beta}\right) + 2  P^{\rho(\mu} \Gamma^{\nu)}_{\sigma(\gamma} \Gamma^{\sigma}_{\delta)\rho} +  P^{\rho\sigma} \Gamma^{\mu}_{\rho(\gamma}\Gamma^{\nu}_{\delta)\sigma} \,.\nonumber\\\label{inverseVar2}
\end{eqnarray}
With the help of the transformation formulae, we can then cast the first covariance equation (\ref{invaronedegP2}) into the form
\begin{equation}\label{invarS}
\frac{\partial C_{B_1\dots B_N}}{\partial S_{\alpha\beta\gamma\delta}}=0\,,
\end{equation}
and the second covariance equation (\ref{invartwodegP2}) can be rewritten in terms of the new arguments as
\begin{equation}\label{invarGamma}
\frac{\partial C_{B_1\dots B_N}}{\partial \Gamma^{\alpha}_{\beta\gamma}}=0\,.
\end{equation}
In other words, the potentials $C_{B_1\dots B_N}$ cannot explicitly depend on the new non-tensorial variables $\Gamma^{\alpha}_{\beta\gamma}$ and $S_{\alpha\beta\gamma\delta}$, but we have that $C_{B_1\dots B_N}=C_{B_1\dots B_N}( P^{\alpha\beta}, R_{\alpha\beta\gamma\delta})$. This is of course what one would expect according to the well-known theorem that the Riemann tensor is the only tensor that can be formed from a metric and its first and second derivatives.

This procedure of changing the arguments on which the potentials depend can be generalized to all cases where, in addition to a metric, one has an arbitrary set of other hypersurface tensor fields $ G^A$. The first and second {\it partial} derivatives of the additional fields $ G^A$ can then be replaced by the first and the symmetrized second {\it covariant} derivatives of $ G^A$ using the torsion-free and metric compatible Levi-Civita connection of the metric at hand. 

For instance, if one has, in addition to $P^{\alpha\beta}$ also scalar and covector hypersurface fields $\phi$ and $g_\alpha$,  the symmetrized covariant derivatives of the fields $ \phi$ and $ g_\alpha$ are given by
\begin{eqnarray}
  \phi_{; \rho \sigma} &=&  \phi_{,\rho \sigma} - \Gamma^{\mu}_{\rho \sigma}  \phi_{,\mu}\,, \label{Var4}\\
  g_{\alpha;\beta}    &=&  g_{\alpha,\beta} -  g_\mu \Gamma^{\mu}_{\alpha \beta}\,, \label{Var5}\\
 g_{\alpha; (\beta \gamma)}  &=&  g_{\alpha,\beta \gamma} - 2\, g_{\mu,(\gamma} \Gamma^{\mu}_{\beta)\alpha} -  g_{\alpha,\mu} \Gamma^{\mu}_{\beta \gamma} \nonumber\\
&&-  g_\mu \left(\Gamma^{\mu}_{\alpha (\beta,\gamma)} - \Gamma^\mu_{\alpha \nu} \Gamma^{\nu}_{\beta \gamma} - \Gamma^{\mu}_{\nu (\beta} \Gamma^{\nu}_{\gamma)\alpha} \right)\,, \label{Var6}
\end{eqnarray}
from which the partial derivatives of the variables $  \phi$ and $ g_\alpha$ are recovered by virtue of  
\begin{eqnarray}
 \phi_{,\rho \sigma}&=& \phi_{;\rho \sigma} + \Gamma^\mu_{\rho\sigma}  \phi_{;\mu}\,,\\
 g_{\alpha,\beta}&=&  g_{\alpha;\beta} + \Gamma^{\mu}_{\alpha\beta}  g_{\mu} \\
 g_{\alpha,\beta\gamma}&=&  g_{\alpha;(\beta\gamma)} +  g_{\mu;\nu}\left[ 2 \Gamma^{\nu}_{\alpha(\gamma}\delta^{\mu}_{\beta)} + \Gamma^{\mu}_{\gamma\beta} \delta^{\nu}_{\alpha}\right] \label{inverseVar4}\\
&& +\frac{1}{6}  g^{\mu} \Big[S_{\mu\alpha\beta\gamma} - R_{\alpha\beta\mu\gamma} - R_{\alpha\gamma\mu\beta} \label{inverseVar5}\\
&&\quad\qquad- 2  P_{\rho\sigma} (\Gamma^{\rho}_{\beta\gamma} \Gamma^{\sigma}_{\alpha\mu}+ \Gamma^{\rho}_{\mu\beta} \Gamma^{\sigma}_{\gamma\alpha} + \Gamma^{\rho}_{\mu\gamma} \Gamma^{\sigma}_{\beta\gamma})\Big]\,.\label{inverseVar6}
\end{eqnarray}

The antisymmetric part of the second covariant derivatives of the fields $ G^A$ does not have to be considered, because it can always be expressed by the Riemann tensor and the undifferentiated fields $ G^A$. After rewriting the respective covariance equations, one again ends up with equations (\ref{invarS}) and (\ref{invarGamma}). In particular, this can be done for all hypersurface point particle geometries of arbitrary degree by formally employing the particular field $P^{\alpha \beta}:=P(\epsilon^\alpha, \epsilon^\beta, n, \dots, n)$ as a metric, and treating all other tensor fields $ P^{\alpha_1\dots\alpha_I}:=P(\epsilon^{\alpha_1}, \dots,\epsilon^{\alpha_I}, n \dots, n)$, for $I=3,\dots,\deg P$, as additional fields. It can also be done for area metric geometry by employing the tensor field $ G^{\alpha\beta}$ as a metric, with respect to which one defines the Levi-Civita connection and the Riemann tensor. However, although we are always guaranteed---by the bi-hyperbolicity and the energy-distinguishing properties---that the tensor field $ P^{\alpha\beta}$, which is distinguished by the matter field equations one employs, can be formally used as a metric tensor on a given hypersurface in $M$, it might not be possible to find an invertible transformation of arguments from $G^A, \partial G^A, \partial\partial G^A, \dots$ to a new set of arguments, which contains $P^{\alpha\beta}$. Nevertheless, if such a transformation exists, one can proceed to rewrite the master equations with respect to these new arguments. 

\section{Explicit mechanism determining the fields $g_\alpha$ in the second case study}
\label{sec:concretemodel}
The most general gravitational dynamics that can underly the predictive and quantizable $SO(p,q)$-violating Dirac dynamics considered in the second case study were found to be unique up to  two freely specifiable functions, $a_1( \phi)$ and $C_{(0)}( \phi,\nabla^{\alpha} \phi\nabla_\alpha \phi,  g^\alpha\nabla_\alpha  \phi, g_\alpha  g^\alpha)$. While the function $a_1(\phi)$ merely mediates the derivative coupling between the metric $P^{\alpha\beta}$ and the scalar field $\phi$, the role of the function $C_{(0)}$ can only be revealed by an explicit solution of the recursion relation derived in Step 9 of the second case study.  Since, apart from the specific set of arguments it depends on, the potential $C_{(0)}$ is completely undetermined by the master equations, we can freely prescribe any additional condition that is compatible with the master equations and at the same time allows to determine $C_{(0)}$. The additional assumption we would like to introduce here, for definiteness, is that the Lagrangian depends at most quadratically on the velocities $K$. Since the most general solution of the recursion relation can be obtained rather straightforwardly under this assumption, a sketch of the derivation shall suffice. First of all, we can ignore the dependence of the functions $C_{(N)}$ on the scalar field $ \phi$ itself. There is no way to constrain this dependence in any way. We simply need to keep in mind that any integration constants, which arise when solving the recursion relations, must be turned into arbitrary functions of $ \phi$ at the end. Introducing the shorthand notations $\Omega=\nabla^{\alpha} \phi\nabla_\alpha \phi$, $\Psi= g_\alpha\nabla^\alpha\phi$ and $\xi= g_\alpha  g^\alpha$ for the arguments of the functions $C_{(N)}$, the general recursion relation takes the form
\begin{equation*}
C_{N+1}=\frac{1}{\Omega}\frac{N!}{(N+1)!}\left[\frac{\partial C_{(N)}}{\partial\Psi}(\Omega+\Psi^2) + 2 \,\frac{\partial C_{(N)}}{\partial \xi}(\Psi+\xi\Psi)+ 2 \Omega\,\frac{\partial C_{N-1}}{\partial \Omega} + \Psi\,\frac{\partial C_{N-1}}{\partial \Psi}\right]\,.
\end{equation*}
Now, assuming that $C_{(N)}=0$ for all $N\geq 3$, we can immediately integrate this equation for $N=3$, which yields
\begin{equation*}
C_{(2)}=A(\xi) \Omega^{2 n}\Psi^{-n} + B(\xi)
\end{equation*}
for some constant $n$ and, up to now, freely specifiable functions $A(\xi)$ and $B(\xi)$. Reinserting this result into the same equation for $N=2$ determines $C_{(1)}$, and reinserting both into the equation for $N=1$ yields $C_{(0)}$. All additional unknown functions, which arise in this process, can then be determined by inserting $C_{(1)}$ and $C_{(0)}$ into the formula for the recursion start. This leads to the condition $n(n+1)(n+2) A(\xi)=0$ and we may then determine all possible solutions for which any of these factors vanish. After a fair amount of algebra, one observes that the cases $A(\xi)=0$, $n=-1$ and $n=-2$ are actually equivalent. Finally, the most general solution for the second part $L_2$ (determined by the recursion formula and that for the recursion start) of the full gravitational Lagrangian, under the condition that it is at most quadratic in the velocities $ K$, is
\begin{eqnarray*}
L_2&=&\sqrt{-\det  P_{\alpha \beta}}\,\Big[\frac{a_3( \phi)}{1+ g_\alpha  g^\alpha}\, K^2+\frac{a_4( \phi)}{(1+ g_\alpha  g^\alpha)^{1/2}}\, K^2+a_5( \phi) \, K^2\nonumber\\
&& +\frac{ 2 a_3( \phi) }{1+ g_\alpha  g^\alpha}\, g^\beta \nabla_\beta  \phi\, K+\frac{a_4( \phi)}{(1+ g_\alpha  g^\alpha)^{1/2}}\, g^\beta \nabla_\beta  \phi\, K+\frac{a_6( \phi)}{(1+ g_\alpha  g^\alpha)^{1/2}}  g^\beta\nabla_\beta \phi\, K\nonumber\\
&&+ \frac{a_3( \phi)}{1+ g_\alpha  g^\alpha} ( g^\beta\nabla_\beta  \phi)^2 - \frac{a_4( \phi)}{(1+ g_\alpha  g^{\alpha})^{1/2}} \nabla^{\beta} \phi\nabla_\beta \phi +\frac{a_6( \phi)}{(1+ g_\alpha  g^\alpha)^{1/2}}  g^\beta \nabla_\beta\phi\nonumber \\
&&+ \,a_5( \phi)\, \nabla^\beta \phi\nabla_\beta \phi+\, a_7( \phi)\, \Big]\,.
\end{eqnarray*}
The last two lines denote the most general form for the potential $C_{(0)}$ that leads to a Lagrangian that is at most quadratic in the scalar velocities $ K$. As we have mentioned already, the free functions $a_3( \phi),\dots,a_7( \phi)$ cannot be further constrained, so that there is a sizable class of possible gravitational theories that can underlie the matter field equations employed in the second case study.

In order to understand the fate of the geometric variable $ g_\alpha$, we first investigate a special case of such a theory. For definiteness, we will specialise to a particularly simple solution for the Lagrangian in order to study the dynamical properties of the derived gravitational theory. We set $a_1( \phi)\equiv-\kappa=\text{const}$, $a_3( \phi)\equiv\mu=\text{const}$, and all other $a_4,\,\dots\,,a_7\equiv0$. Then the Lagrangian reads
\begin{equation*}
L=\sqrt{-\det  P_{\alpha \beta}} \left[\kappa C_{\alpha \beta \gamma \delta} K^{\alpha \beta} K^{\gamma \delta} - \kappa R + \mu \frac{ K^2}{1+ g_\alpha  g^\alpha} + 2 \mu \frac{ K}{1+ g_\alpha  g^\alpha} + \mu \frac{( g^\beta\nabla_\beta \phi)^2}{1+ g_\alpha  g^\alpha}\right]\,.
\end{equation*}
It is easy to analyse the dynamics of this theory in the canonical spacetime picture. To this end, one performs the inverse Legendre transformation of the above Lagrangian with respect to the velocities $ K^A$. Since the Lagrangian is singular in the velocity $ K_\alpha$, one picks up  additional Lagrange multipliers $\Lambda_\alpha$ in the process. After performing the Legendre transformation, the complete Hamiltonian for our particular gravity theory becomes
\begin{eqnarray*}
 H &=& \int_\Sigma dy\, \Big[ N(y) \Big\{ \frac{1}{4 \kappa \sqrt{-  P}} C^{\alpha \beta \gamma \delta} \pi_{\alpha \beta}  \pi_{\gamma \delta} + \kappa \sqrt{- P} R+ \frac{1}{4\mu \sqrt{- P}}  \pi^2 (1+ g_\alpha  g^\alpha)  \nonumber\\
&&- \pi  g^\alpha\nabla_\alpha\phi  -\sqrt{-  P}\,\frac{(\mu-1)^2}{\mu} \frac{( g^{\beta}\nabla_\beta\phi)^2}{(1+ g_\alpha  g^\alpha)} + \Lambda_\alpha  \pi^\alpha- \partial_\gamma ( \pi^\gamma+ g^\gamma  g_\alpha  \pi^\alpha)\Big\}(y)\nonumber\\
&&+\left\{ \pi_{\alpha \beta}\,\mathcal L_{\vec N}  P^{\alpha \beta}+ \pi\, \mathcal L_{\vec N}  \phi +  \pi^{\alpha} \,\mathcal L_{\vec N}  g_\alpha\right\}(y)  \Big] \,,
\end{eqnarray*}
with the potential $C^{\alpha\beta\gamma \delta}=4  P^{\alpha (\gamma} P^{ \delta) \beta} - 2  P^{\alpha \beta}  P^{\gamma \delta}$, and we used the shorthand $\sqrt{- P}:=\sqrt{-\det  P_{\alpha\beta}}$. For further analysis, we simplify matters by setting $\mu=1$. The Lagrange multiplier $\Lambda_\alpha$ enforces $\pi^\alpha(y)\equiv0$ as an additional constraint. Since $\pi^\alpha(y)\equiv0$ has to hold for all values of the evolution parameter $t$, this also implies that $\dot \pi^\alpha(y)=0$. However, Hamilton's equations for the variable $ g_\alpha$ using the above Hamiltonian yield
\begin{equation*}
\dot{ \pi}^\alpha(y)\approx - N(y) \left[\frac{1}{2\sqrt{- P}} \, \pi^2  g^\alpha -  \pi \nabla^\alpha  \phi\right](y)\,,
\end{equation*}
where the weak equality `$\approx$' means that we already made use of the constraint $ \pi^\alpha=0$. Hence, the variable $ g_\alpha$ is completely determined by the solutions of the equations of motion for the scalar field $ \phi$ and the metric $ P^{\alpha \beta}$ by
\begin{equation*}
  g_\alpha(y)=2 \left[\sqrt{- P}\, \frac{\nabla_\alpha  \phi}{ \pi}\,\right](y).
\end{equation*}
Hamilton's equations for the variable $ \pi^\alpha$ can be used to determine the Lagrange multiplier $\Lambda_\alpha$, and to eliminate the variable $g_\alpha$ and the momentum $\pi^{\alpha}$ altogether. From the remaining equations of motion, it can then be checked that the effective Hamiltonian for the dynamics of the scalar field $ \phi$ and the metric $ P^{\alpha\beta}$ is given by
\begin{eqnarray*}
 H&=& \int_\Sigma dy\, \Big[ N\, \Big\{ \frac{1}{4 \kappa \sqrt-  P} C^{\alpha \beta \gamma \delta} \pi_{\alpha \beta}  \pi_{\gamma \delta} + \kappa \sqrt{- P} R+ \frac{1}{4\sqrt{- P}} \pi^2 -\sqrt{- P}\nabla^{\alpha} \phi\nabla_\alpha  \phi\Big\}\nonumber\\
&&\quad\qquad+\left\{ \pi_{\alpha \beta}\,\mathcal L_{\vec N}  P^{\alpha \beta}+ \pi\, \mathcal L_{\vec N}  \phi \right\}  \Big](y)\,,
\end{eqnarray*}
which is mathematically equivalent to a massless scalar field non-derivatively coupled to Einstein gravity.

Our considerations show that, although the variable $ g_\alpha$ is not dynamical in the sense that it satisfies its own dynamical equations of motion, it is nevertheless completely determined by the dynamics of the other degrees of freedom of the theory. 


\end{document}